\documentclass[12pt]{iopart}
\usepackage{graphicx}
\usepackage{amssymb}
\usepackage{amsfonts}
\begin{document}

\title[Crystallization of magnetic dipolar monolayers]
{Crystallization of magnetic dipolar monolayers: a density functional approach}

\author{Sven~van Teeffelen\footnote{Corresponding author. E-mail: teeffelen@thphy.uni-duesseldorf.de}, Hartmut~L{\"o}wen, and Christos~N.~Likos}

\address
{Institut f{\"u}r Theoretische Physik II: Weiche Materie,\\
Heinrich-Heine-Universit{\"a}t D{\"u}sseldorf,
Universit{\"a}tsstra{\ss}e 1, D-40225 D{\"u}sseldorf, Germany}

\begin{abstract}
  We employ density functional theory to study in detail the
  crystallization of super-paramagnetic particles in two dimensions
  under the influence of an external magnetic field that lies
  perpendicular to the confining plane.  The field induces
  non-fluctuating magnetic dipoles on the particles, resulting into an
  interparticle interaction that scales as the inverse cube of the
  distance separating them. In line with previous findings for
  long-range interactions in three spatial dimensions, we find that
  explicit inclusion of liquid-state structural information on the
  {\it triplet} correlations is crucial to yield theoretical
  predictions that agree quantitatively with experiment.  A
  non-perturbative treatment is superior to the oft-employed
  functional Taylor expansions, truncated at second or third order.
  We go beyond the usual Gaussian parametrization of the density
  site-orbitals by performing free minimizations with respect to both
  the shape and the normalization of the profiles, allowing for finite
  defect concentrations.
\end{abstract}

\pacs{64.10.+h,64.70.Dv,82.70.Dd}
\maketitle
%
\section{Introduction}

Classical density functional theory (DFT) is the method of choice to
the study of inhomogeneous fluids \cite{Evans:79}. Perhaps the most
extreme inhomogeneities arise in a crystalline solid, where the
density field $\rho({\bi r})$ is both periodic and shows extreme
differences between its local values on the lattice sites and in the
interstitial regions. DFT has been successfully applied to the problem
of crystallization of a number of different
systems~\cite{Oxtoby:91,Singh:91,Loewen:94,Loewen:02,Likos:01}, {\it
  mostly} in three spatial dimensions. Here, the most popular system
is the prototype of hard spheres, for which a geometry-based
theory~\cite{Rosenfeld:89,Rosenfeld:97,Schmidt:03} has proven quite
successful. For soft
interactions~\cite{Curtin:88,Curtin:85,deKuijper:90}, however, where
one cannot assign geometrical measures to the interacting
point-particles, one has to resort to other functionals. In
particular, it has been shown \cite{Likos:92} that for long-range
interactions, structural information of the liquid on the pair-level
is insufficient and triplet fluid correlations should be allowed to
explicitly flow into the construction of the functional.  Even less is
known for crystallization in two spatial
dimensions~\cite{Zeng:90,Rosenfeld:90,Tejero:93,Teeffelen:06}.  Here,
we consider a combination of the two above-mentioned cases in
considering {\it long range interactions} in {\it two spatial
  dimensions} and we study in detail the role played by accurate
liquid-state information on triplet correlations in determining phase
boundaries between a fluid and the coexisting crystal.

In this paper, we study freezing of a classical two-dimensional model
fluid, namely of a fluid of aligned dipoles directed perpendicular to
the 2D-plane and repelling each other with a soft $1/r^3$
inverse-power pair potential, with the help of density functional
theory (DFT).  In Ref.~\cite{Teeffelen:06} we studied freezing of the
the dipolar system with the modified weighted density approximation
and its extension to third order correlation functions.  Within this
paper we will extend our previous study in several ways: We allow for
a finite defect concentration and relax the constraint of Gaussian
density peaks in the crystalline phase, as, e.g., suggested for hard
sphere crystals in Ref.~\cite{Ohnesorge:93}.  Furthermore, we
systematically study the influence of perturbative and
non-perturbative inclusion of higher order correlation functions of
the liquid in the density functional approximation on the freezing
transition. We employ two different approximations to the
three-particle correlation functions, which lead to substantially
different results, therefore signalling the importance of an accurate
approximation of the latter.

We use different approximations to the DFT---based on the famous and
powerful approach by Ramakrishnan and Yussouff~\cite{Ramakrishnan:79},
but extending on the latter in taking higher-order terms into account,
as will be described below. The quantity to be approximated in the DFT
of freezing is the excess Helmholtz free energy functional $F_{\rm
  ex}[\rho({\bi r})]$, a unique functional of the inhomogeneous
one-particle density $\rho({\bi r})$ of the solid~\cite{Evans:79}. The
uniqueness property implies that the excess free energy can be
formally expanded about the excess free energy of a homogeneous fluid
at a uniform density $\rho$ in terms of density difference
$\Delta\rho({\bi r})=\rho({\bi r})-\rho$:
\begin{eqnarray}
\fl  \beta F_{\rm ex}[\rho({\bi r})]=\beta F_{\rm ex}(\rho)-
\sum_{n=1}^\infty \frac{1}{n!}\int_V\,\mathrm{d}  \bi r_1 \dots \mathrm{d}  \bi r_n 
  c_0^{(n)}\left(\bi r,\dots,\bi r_n; \rho\right)
  \Delta\rho(\bi r_1)\dots\Delta\rho(\bi r_n),\label{eq:Fex_expansion}
\end{eqnarray}
where $\beta=1/(k_BT)$ and $V$ is the volume occupied of the system.
$F_{\rm ex}(\rho)$ is the Helmholtz excess free energy and the
$c_0^{(n)}$ are the $n$-particle direct correlation functions of the
fluid, which are well known up to second order for dipolar
fluids~\cite{Teeffelen:06}.

Within the theory of Ramakrishnan and Yussouff this series expansion is
truncated at second order. We therefore refer to the theory as
``second order theory'' (SOT). Part of the reason for this truncation
lies in
poor knowledge about higher than second order correlation functions;
the truncation is not well justified in the problem of freezing, since here
$\Delta \rho$ is not a small parameter. In particular, it has been
extensively shown that the SOT fails to accurately predict freezing
for systems interacting via long-range pair potentials for
three-dimensional systems~\cite{Likos:92,Laird:90}. We will show in
this work, that also for the two-dimensional dipolar system the SOT
highly underestimates the stability of the crystal.  Therefore,
several approaches have been employed to include higher than second-order
terms in the expansion---in a perturbative~\cite{Curtin:88} or
non-perturbative way~\cite{Curtin:85,Likos:92,Denton:89b,Likos:93}.

The simplest attempt to go beyond the SOT is to explicitly include the
third order term in the expansion in
equation~(\ref{eq:Fex_expansion}), which we refer to as ``third order
theory'' (TOT). Employing the TOT demands an approximate form of the
three-particle direct correlation function $c_0^{(3)}(\bi r,\bi
r^\prime;\rho)$ of the fluid.  We will show here, that---given an
accurate expression for $c_0^{(3)}(\bi r,\bi
r^\prime;\rho)$---including this term substantially improves the
predicted freezing temperature of the long-range $1/r^3$-fluid (in
line with previous findings for long-range interactions in
3D~\cite{Barrat:88}).

A third approach to the DFT we follow here, is the Modified
Weighted-Density Approximation (MWDA)~\cite{Denton:89b} by Denton and
Ashcroft which we already presented for the dipolar system in brief in
a previous paper~\cite{Teeffelen:06}. This approach includes first
and second order correlation functions of the fluid {\it exactly} (as
in the SOT) and higher order correlation functions in a
non-perturbative, implicit fashion. We find that the MWDA, in two
dimensions, slightly shifts the freezing transition to higher
temperature as compared to the SOT, still highly underestimating the
stability of the solid state. In a fourth approach we employ the so
called ``extended modified weighted-density approximation'' (EMA), as
suggested in Refs.~\cite{Likos:92,Likos:93}. Different from the MWDA,
this approximation to the density functional now includes not only
first and second, but also third order correlation functions of the
fluid {\it exactly} (as in the TOT). Higher than third-order
correlation functions are contained in a non-perturbative, implicit
fashion, following a similar scheme as in the MWDA. For the dipolar
system we find that this approach leads to a very accurate value of
the freezing transition temperature, lying slightly above the one
obtained from the simpler TOT.  The two-particle correlation functions
of the liquid are obtained from liquid state integral equation theory and
from simulation. The three-particle correlation functions are obtained
applying two approximations, both based on the two-particle
correlation functions: The first approximation used is by Denton and
Ashcroft (DA)~\cite{Denton:89}, and the second is by Barrat, Hansen,
and Pastore (BHP)~\cite{Barrat:88b}.

We find that the inclusion of higher order correlation functions in a
perturbative (TOT) or non-perturbative (EMA) way subsequently increase
the freezing transition temperature, thus broadening the range of the
thermodynamical stability of the crystal. In fact, we find the
freezing transition temperature to be in good agreement with
experiment~\cite{Zahn:99} and
simulation~\cite{Lin:06,Haghgooie:05,Loewen:96}. The importance of
inclusion of third order correlation functions is addressed to the
long-range nature of the dipole-dipole pair interaction.

The rest of this work is organized as follows: In section~2 we give a
brief description of the MWDA and of the EMA. In section~3 we apply
the different approximations to the DFT to freezing of monodispers
two-dimensional liquids. The theory is adapted to the dipolar system
under study in section~4. In section~5 we present the resulting phase
diagrams and different structural properties of the crystalline
system, and we conclude in section~6.
%
\section{Modified weighted-density approximation and its extension
  to third-order correlation functions}
\label{sec:mwdaema}
It is well known that the intrinsic Helmholtz free energy of an
inhomogeneous system can be divided into an ``ideal'' and an ``excess''
part,
\begin{equation}\label{eq:ftot}
F\left[\rho(\bi r)\right]=
F_{\rm{id}}\left[\rho(\bi r)\right]+F_{\rm{ex}}\left[\rho(\bi r)\right]\,.
\end{equation}
The ``ideal'' term 
\begin{equation}\label{eq:fid}
F_{\rm{id}}\left[\rho(\bi r)\right]= 
\beta^{-1}\int\,\mathrm{d} \bi r \rho(\bi r)\left\{
\ln\left[\rho(\bi r)\Lambda^2\right]-1\right\}\,,
\end{equation}
is known exactly. In equation~(\ref{eq:fid}) $\Lambda$ is the
thermal de Broglie wavelength. The excess part can only be calculated
approximately. In contrast to the SOT and TOT, within the MWDA and EMA
the excess free energy of the inhomogeneous system is approximated by
setting it equal to the excess free energy of a uniform liquid
evaluated at a weighted density $\hat{\rho}$,
\begin{equation}\label{eq:fmwda}F_{\rm{ex}}\left[\rho(\bi r)\right]\approx
F_{\rm{ex}}^{\rm M/E}\left[\rho(\bi r)\right]=  N f_0(\hat\rho^{\rm M/E})\,,
\end{equation}
where superscripts denote the approximations to the DFT, MWDA (M) and
EMA (E), respectively. $N$ is the number of particles in the system
and $f_0(\hat\rho)$ is the excess free energy per particle of the
liquid at the weighted density $\hat\rho$. The latter is expressed as
\begin{eqnarray}\fl \hat{\rho}^{\rm M/E}\left[\rho (\bi r)\right]=
   \frac{1}{N}\int\,\mathrm{d} \bi r \, \mathrm{d}
    \bi r^\prime \rho({\bi r})\rho({\bi r}^\prime) w\left({\bi r}-{\bi
      r}^\prime;\hat{\rho}\right) \nonumber\\
+  \frac{1}{N^2}\int\,\mathrm{d} \bi r \, \mathrm{d} \bi r^\prime \, \mathrm{d} \bi r^{\prime\prime}
  \rho({\bi  r})\rho({\bi r}^\prime)\rho({\bi r}^{\prime\prime}) v\left({\bi
      r}-{\bi r}^\prime,{\bi r}-{\bi
      r}^{\prime\prime};\hat{\rho}\right)\,,\label{eq:rhohat}
\end{eqnarray}
where the second term only appears in the EMA and not in the MWDA.
The weight functions $w(\bi{r};\rho)$ and $v(\bi r,\bi r^\prime;\rho)$
are determined in such a way that the approximate functional
$F_{\rm{ex}}^{\rm M/E}\left[\rho(\bi r)\right]$ is exact up to second
(MWDA) or third (EMA) order in density difference $\Delta\rho(\bi
r)=(\rho(\bi r)-\rho)$, i.e., up to that order equation~(\ref{eq:fmwda})
and equation~(\ref{eq:Fex_expansion}) do agree.  Note that the weighted
density $\hat\rho$ is determined self-consistently, as it appears as
an argument of both weight functions.  In order to obtain equality
with equation~(\ref{eq:Fex_expansion}) up to second or third order in
$\Delta\rho$ we demand the weight functions to be normalized, i.e.,
\begin{equation}
  \label{eq:normalization}
\int_V\,\mathrm{d} \bi r  w(\bi{r};\rho)+
\frac{1}{V}\int_V\,\mathrm{d} \bi r  \int_V\,\mathrm{d} \bi r' 
v(\bi r,\bi r^\prime;\rho)=1\,.
\end{equation}
and to fulfill the requirements
\begin{eqnarray}
\lim_{\rho(\bi r)\rightarrow\rho} \left[\frac{\delta^2 F_{\rm{ex}}^{\rm{M/E}}}{\delta \rho(\bi r)\delta \rho(\bi
  r^\prime)}\right]&=-\beta^{-1}c_0^{(2)}\left(\bi r - \bi r^\prime; \rho\right)\,,\nonumber\\
\lim_{\rho(\bi r)\rightarrow\rho} \left[\frac{\delta^3 F_{\rm{ex}}^{\rm{E}}}{\delta \rho(\bi r)\delta \rho(\bi
  r^\prime)\delta \rho(\bi
  r^{\prime\prime})}\right]&=-\beta^{-1}c_0^{(3)}\left(\bi r - \bi
r^\prime, \bi r - \bi r^{\prime\prime};\rho\right)\,,\label{eq:deltafexnachdeltarho}
\end{eqnarray}
where $c_0^{(2)}(\bi r; \rho)$ and $c_0^{(3)}(\bi r, \bi
r^{\prime};\rho)$ are the two- and three-particle correlation
functions of the liquid with density $\rho$ which are an input to the
theory.  These conditions uniquely determine the weight functions. In
order to obtain the simple algebraic equations for $v$ and $w$ that
can be found in Ref.~\cite{Likos:93} a further approximation has to be
made: The inner integral in the second term of
equation~(\ref{eq:normalization}) is assumed to be equal to a
constant, $C$ (demanding the first term in
equation~(\ref{eq:normalization}) to be equal to $1-C$), where $C$ is
independent of the fixed space coordinate of the weight function
$v(\bi r,\bi r^\prime;\rho)$. The weighted density $\hat\rho$ in
equation~(\ref{eq:rhohat}) is independent of the choice of
$C$~\cite{Likos:93}.

For non-zero wave vectors (${\bi k} \neq 0,{\bi k}^\prime \neq 0$, or
${\bi k} + {\bi k}^\prime \neq 0$), the Fourier transforms of the
weight functions $\tilde w(k; \rho)$ and $\tilde v(\bi k, \bi
k^\prime; \rho)$ are simply proportional to the Fourier transforms of
the second- and third-order direct correlation functions $\tilde
c_0^{(2)}(k;\rho)$ and $\tilde c_0^{(3)}({\bi k},{\bi k}^\prime;
\rho)$, respectively:
\begin{eqnarray}
-\beta^{-1}\tilde c_0^{(2)}(k;\rho)&=2f_0^\prime(\rho)\tilde w(k; \rho)\,,\nonumber\\
-\beta^{-1}\tilde c_0^{(3)}({\bi k},{\bi k}^\prime;\rho)&=
6 f_0^\prime(\rho)\tilde v({\bi k},{\bi k}^\prime; \rho)\,.\label{eq:wvcorr}
\end{eqnarray}
Furthermore, equation~(\ref{eq:rhohat}), together with
equations~(\ref{eq:normalization}) and (\ref{eq:deltafexnachdeltarho}) guarantee
fulfillment of the sum rules
\begin{eqnarray}
\beta^{-1}\tilde c_0^{(2)\rm M/E}(k=0;\rho)&=2f_0^\prime(\rho)+\rho f_0^{\prime\prime}(\rho)\,,\nonumber\\
\tilde c_0^{(3)\rm M/E}({\bi k},{\bi k}^\prime=0; \rho)&=
\tilde c_0^{(3)\rm M/E}({\bi k},- {\bi k}; \rho)=\frac{\partial \tilde c_0^{(2)}(k;\rho)}{\partial \rho}\,,\label{eq:sumrules}
\end{eqnarray}
where the former is the compressibility sum rule, and where the
superscripts on the correlation functions indicate that these
functions are the Fourier transforms of the functional derivatives of
the approximate excess free energy functionals in the limit of
constant average density $\rho$ [c.f. equation~(\ref{eq:deltafexnachdeltarho})].
The primes on the excess free energy density $f_0$ denote derivatives
with respect to  density.

Due to the self-consistency requirement, the approximate excess free
energies of both the MWDA and the EMA include contributions from
arbitrarily many higher orders. However, if expanded about the excess
free energy of a fluid with the same average density as the
inhomogeneous system according to equation~(\ref{eq:Fex_expansion}), the
MWDA only gives even order terms and estimates the odd order terms
zero. Contrary, the EMA includes, approximately, contributions from
all higher order terms. In particular, it includes the exact
third-order term, which is an input to the theory.
%
\section{Application of the different approximations to the DFT to
  freezing of monodisperse two-dimensional liquids}
\label{sec:parametrization}
In order to find the equilibrium one-particle density
$\rho_{\rm{eq}}(\bi r)$ of a system at a given average density $\rho$
and temperature $T$ we minimize the approximate total free energy
functional $F[\rho(\bi r)]$ of equation~(\ref{eq:ftot}) with respect to the
inhomogeneous one-particle density $\rho({\bi r})$ for fixed $\rho$. As
described, for example, in Refs.~\cite{Denton:89b,Likos:93} this
minimization is pursued in a number of subsequent steps, depending on
the kind of approximation: For all approximations to the DFT, first,
an appropriate parametrization for the inhomogeneous one-particle
density is made (we will employ a free minimization in
section~\ref{sec:freemin}). Within the SOT and TOT, we can now, in a
second step, calculate the excess and ideal parts of the Helmholtz
free energy according to equations~(\ref{eq:Fex_expansion}) and (\ref{eq:fid}).
However, within the MWDA and EMA, the excess part is given by
equation~(\ref{eq:fmwda}), with the weighted density $\hat\rho$ obtained in an
intermediate step according to equation~(\ref{eq:rhohat}).  In a final
step, minimization is carried out with respect to all free variables
in the parametrization of $\rho(\bi r)$.

The crystalline one-particle density which we expect to be in
equilibrium for low temperature and/or high density has the symmetry
of the triangular crystal---the quadratic lattice is
thermodynamically unstable for the whole range of accessible
densities/coupling constants and we expect mechanical instability with
respect to the triangular lattice for any coupling. We can therefore
express $\rho({\bi r})$ as a sum over reciprocal lattice vectors (RLV's) of the
triangular lattice:
\begin{equation} \label{eq:rhok}
\rho({\bi r})=\rho
\left[1+\sum_{{\bi K}\neq 0}\mu_{\bi K}e^{i{\bi Kr}}\right]\,,
\end{equation}
where $\rho$ is the average density of the solid, $\{{\bi K}\}$ is the
set of reciprocal lattice vectors (RLV's), and where the $\mu_{\bi K}$
are the dimensionless Fourier components. In terms of Fourier
components the excess part to the Helmholtz free energy within SOT and
TOT now reads
\begin{eqnarray} \label{eq:fexRY_rhok}
\fl  \beta F_{\rm ex}^{\rm S/T}[\rho({\bi r})]/N=\beta f_0(\rho)-
\frac{\rho}{2}\sum_{{\bi K}\neq 0}\mu_{\bi K}^2 \tilde c_0^{(2)}(k;\rho)\nonumber \\
-\frac{\rho^2}{6} \sum_{{\bi K}\neq 0}\sum_{{\bi K^\prime}\neq 0,-{\bi K}}\mu_{\bi K}
\mu_{\bi K^\prime}\mu_{-({\bi K}+{\bi K}^\prime)}\tilde c_0^{(3)}(\bi K,\bi K';\rho)\,,
\end{eqnarray}
the superscript referring to the SOT (S) and to the TOT (T),
respectively. The third term only appears in the TOT.

Within the MWDA and EMA, the weighted density, equation~(\ref{eq:rhohat}),
now reads
\begin{eqnarray} \label{eq:rhohatk}
\fl \hat\rho^{\rm M/E}=\rho\Bigg\{1+\sum_{{\bi K}\neq 0}
\mu_{\bi K}^2 w\left(K;\hat\rho\right)
+\rho\sum_{{\bi K}\neq 0}\sum_{{\bi K^\prime}\neq 0,-{\bi K}}\mu_{\bi K}
\mu_{\bi K^\prime}\mu_{-({\bi K}+{\bi K}^\prime)}
\left[\frac{v\left({\bi K},{\bi K}^\prime;\hat\rho\right)}{N}\right]\Bigg\}\,.
\end{eqnarray}
As in equation~(\ref{eq:rhohat}) the three-particle term only appears in
the EMA.

Since a free minimization of the approximate Helmholtz free energy
with respect to an infinite number of Fourier components $\mu_{\bi K}$
at all RLV's is intractable, we make a simple ansatz for the
one-particle density which is a superposition of normalized Gaussians
centered around the lattice sites of the triangular lattice:
\begin{equation}\label{eq:ansatzrhoinhomogeneous}
\rho(\bi r)= \frac{n_c\alpha}{\pi}\sum_{\bi R}
\exp\left[-\alpha\left|\bi r - \bi R \right|^2\right]\,,
\end{equation}
where $\alpha$ is the localization strength, $n_c$ is the average
number of particles occupying a lattice site, yielding a vacancy
concentration $n_v=1-n_c$, and $\{\bi R\}$ is the set of Bravais
lattice vectors of the triangular lattice with lattice constant
$a=(\sqrt{3}n_c/2\rho)^{1/2}$. Thus, the Fourier components
$\mu_{\bi K}$ now simply read
\begin{equation}\label{eq:fouriercoeff}\mu_{\bi K}=e^{-\bi K^2/4\alpha}\,.
\end{equation}
The ansatz, equation~(\ref{eq:ansatzrhoinhomogeneous}), was chosen in such a
way that the system forms a triangular lattice for any finite $\alpha$
keeping its average density $\rho$ fixed. For $\alpha\rightarrow 0 $
the density profile becomes flat and the system turns into a liquid.
We thus end up with two minimization parameters $\alpha$ and $n_c$.

This ansatz disregards a possible partition of the system into
coexisting liquid and crystal phases of different densities keeping
the overall average density fixed. However, this is accounted for by
performing a common-tangent construction to the crystal and liquid
volume free energy densities in the end.  Furthermore,
equation~(\ref{eq:ansatzrhoinhomogeneous}) disregards the spatial
anisotropy of the density site profile at each lattice site. We will
see in section~\ref{sec:freemin}, where we relax the constraint on the
density peaks, that both, the assumption of isotropy and the Gaussian
shape are well justified close to the positions of the Bravais lattice
vectors, i.e., where the density is reasonably large ($\rho(r)\gtrsim
\rho$).

\begin{figure}
\begin{center}
  \includegraphics[width=7.5cm]{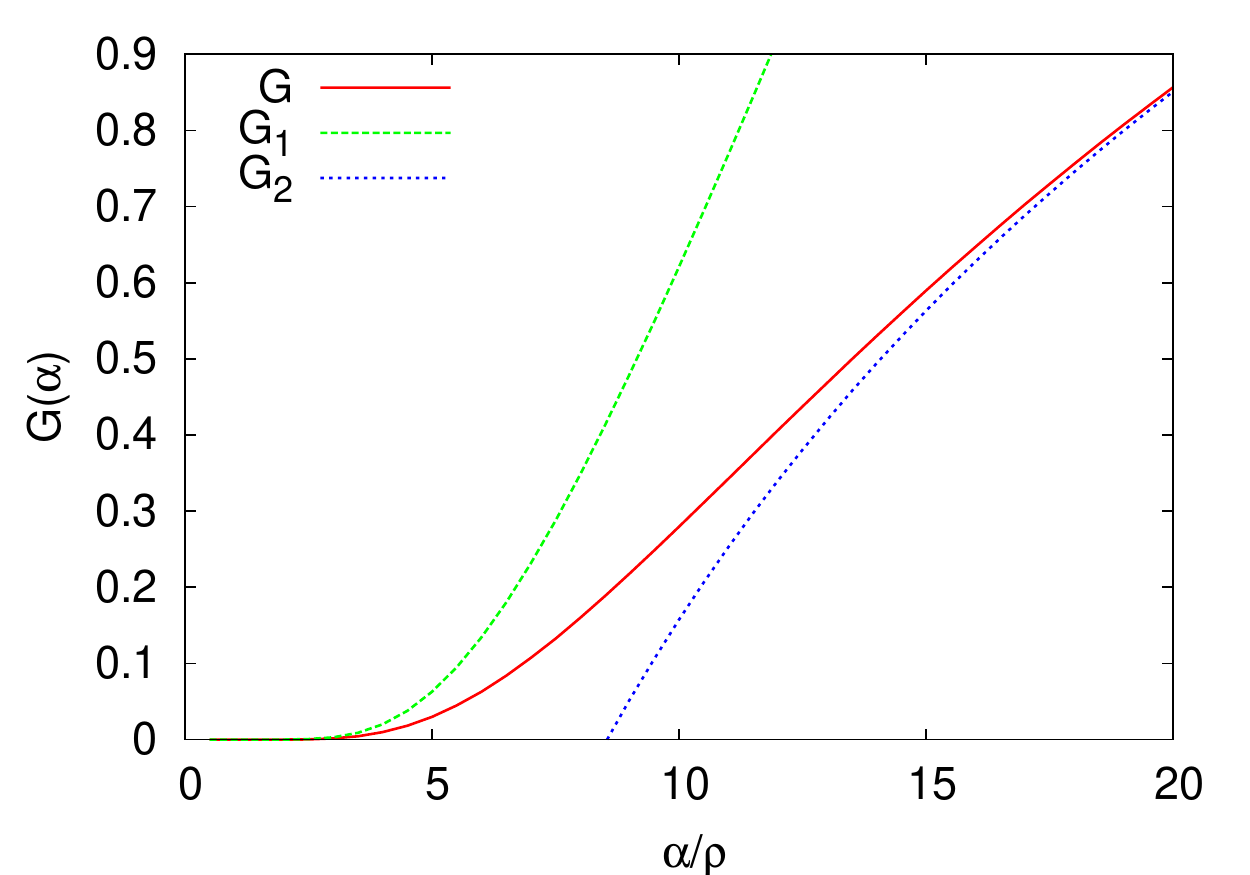}
  \caption{The function $G(\alpha/\rho)$ and its analytically known
    asymptotics for small and large localization strength.}
  \label{fig:fid_comp}
\end{center}\end{figure}

Employing the ansatz of equation~(\ref{eq:ansatzrhoinhomogeneous}) for the
inhomogeneous density, the ideal part of the Helmholtz free energy
[equation~(\ref{eq:fid})] can now be written as a function of $\alpha$ and
$n_c$ only: $F_{\rm{id}}\left[\rho(\bi
  r)\right]=F_{\rm{id}}(\alpha,n_c;\rho)$. For $n_c=1$ it reads
\begin{eqnarray}\label{eq:fidnum}
&\frac{\beta}{N} F_{\rm{id}}(\alpha,n_c=1;\rho)
={\rm const}+\ln(\rho L^2)+G(\alpha^*), \\
&G(\alpha^*)=\int_{A_1}\,\mathrm{d} \bi x 
\frac{\rho(\bi x,\alpha^*,n_c=1)}{\rho}
\ln\left[\frac{\rho(\bi x,\alpha^*,n_c=1)}{\rho}\right]\,,
\end{eqnarray}
where $\rm const$ is an irrelevant constant and $L$ is a
density-independent length scale of the system.  $\bi x=\bi
r\rho^{1/2}$ and $\alpha^*=\alpha/\rho$ are the dimensionless space
coordinate and localization strength, respectively, and the integral
is performed over the area $A_1$ of a unit cell. The function
$G(\alpha^*)$ is approximated for small and large localization
strengths by its analytically known asymptotics
\begin{equation}\label{eq:G}
G(\alpha^*)\simeq\left\{\begin{array}{ll}
G_1(\alpha^*)=\sum_{\bi  K_*\neq0}\exp\left[-{\bi K_*}^2/2\alpha^*\right],&\alpha^*\ll 1\\
G_2(\alpha^*)=\ln(\alpha^*/\pi)-1,&\alpha^*\gg1\,,
\end{array}\right.
\end{equation}
where ${\bi K_*}={\bi K}/\rho$ are the dimensionless RLV's. For
intermediate values of $2\leq\alpha^*\leq50$ the function
$G(\alpha^*)$ was calculated numerically. The function $G(\alpha^*)$
and the asymptotics of equation~(\ref{eq:G}) are plotted as a function
of $\alpha^*$ in figure~\ref{fig:fid_comp}.

The ideal free energy for values $n_c\neq 1$ is obtained via the
simple scaling relation
\begin{equation}
\frac{\beta}{N} F_{\rm{id}}(\alpha,n_c,\rho)=
{\rm const}+\ln(\rho L^2)+G(n_c\alpha^*)\,.
\end{equation}
%
%
\section{The dipolar system}
\label{sec:dipoles}
We now turn to the system of monodisperse particles which repel each
other with an inverse-power pair potential $u(r)=u_0/r^3$, where $u_0$
is a parameter with dimensions of energy $\times$ volume. For the
specific realization of two-dimensional paramagnetic colloids of
susceptibility $\chi$ exposed to a magnetic field ${\bi B}$ which is
directed perpendicular to the 2D plane, we have $u_0 = (\chi {\bi
  B})^2/2$ in Gaussian units~\cite{Froltsov:03}. Here, we assume
perfect alignment of the magnetic dipoles with the external field
which is well justified for $\chi B^2\gg k_BT$~\cite{Zahn:97}. The
thermodynamics and structure depend, due to simple scaling, only on
one relevant dimensionless coupling parameter~\cite{Weeks:81}
\begin{equation}\label{eq:Gammarho}
  \Gamma =  \frac{u_0\rho^{3/2}}{k_BT}\,.
\end{equation}
Therefore, it is convenient to express all quantities in terms of
$\Gamma$ and consider coupling parameters rather than densities via
this scaling relation. Correspondingly, the excess free energy within
the SOT and TOT [equation~\ref{eq:Fex_expansion}] now read
\begin{eqnarray}  
\fl  \beta F_{\rm ex}^{\rm S/T}(\Gamma,\alpha)/N
=\beta f_0(\Gamma)-
\frac{1}{2}\sum_{{\bi K}\neq 0}e^{-K^2/4\alpha} \hat c_0^{(2)}(K;\Gamma) \nonumber \\
-\frac{1}{6} \sum_{{\bi K}\neq 0}\sum_{{\bi K^\prime}\neq 0,-{\bi K}}e^{-(K^2+K'^2+({\bi K}+{\bi K}^\prime)^2)/4\alpha}
\hat c_0^{(3)}(\bi K,\bi K';\Gamma)\,,\label{eq:fexRYgamma}
\end{eqnarray}
the third term only appearing in the TOT. Here, $\Gamma$ is the
coupling constant corresponding to the average density $\rho$
according to equation~(\ref{eq:Gammarho}), $\hat c_0^{(2)}=\rho \tilde
c_0^{(2)}$, and $\hat c_0^{(3)}=\rho^2 \tilde c_0^{(3)}$ are the
dimensionless correlation functions of the fluid in reciprocal space,
respectively.

For the MWDA and EMA, the weighted coupling constants $\hat\Gamma$ now
read
\begin{eqnarray}  
\fl  \hat\Gamma(\Gamma,\alpha)
  =\Gamma\Bigg[1-\frac{1}{3\beta \hat\Gamma f_0^\prime(\hat\Gamma)}\sum_{{\bi K}\neq
      0}e^{-K^2/2\alpha} \hat c_0^{(2)}(K;\hat\Gamma)\nonumber\\    
-\frac{\Gamma  ^{2/3}}{9\beta\hat\Gamma^{5/3} f_0^\prime(\hat\Gamma)}
\sum_{{\bi K}\neq 0}
    \sum_{{\bi K}^\prime\neq 0,-{\bi K}}e^{-(K^2+K^{\prime2}+\bi{K K}^\prime)/2\alpha} 
    \hat  c_0^{(3)}({\bi K},{\bi K}^\prime;\hat\Gamma)\Bigg]^{3/2}\,,\label{eq:gammahatrhohat}
\end{eqnarray}
where $f_0^\prime(\Gamma)$ is the derivative of the excess free energy
density with respect to coupling constant. As in
equation~(\ref{eq:rhohatk}) the third term only appears in the EMA.

In order to solve equations~(\ref{eq:fexRYgamma}) and
(\ref{eq:gammahatrhohat}) we need the two- and three-particle
correlation functions $\hat c_0^{(2)}(k;\Gamma)$ and $\hat
c_0^{(3)}(k;\Gamma)$ and the excess free energy density $f_0(\Gamma)$
of the corresponding liquid for a wide range of coupling constants
$\Gamma$. The two-particle correlation function is obtained with
liquid state integral equation theory or from computer simulations. In
the first case, following the procedure described
in Ref.~\cite{Hoffmann:06} we solve the Ornstein-Zernicke (OZ)
equation~\cite{hansen-mcdonald:86}
\begin{equation}\label{eq:hq}
  \hat h(k) = \frac{\hat c_0^{(2)}(k)}{1 - \hat c_0^{(2)}(k)}\,,
\end{equation}
which relates the dimensionless Fourier transform $\hat h(k)=\rho
\tilde h(k)$ of the total correlation function $h(r)$ to the direct
pair correlation function $\hat c_0^{(2)}(k)$, numerically. Note that
the density has been absorbed in both the Fourier transform of the
total correlation function $\hat h(k)$ and in the direct correlation
function $\hat c_0^{(2)}(k)$.  The total correlation function
is connected to the pair distribution function via $g(r)=h(r)+1$.

The solution of equation~(\ref{eq:hq}) for the two unknown quantities
$\hat h(k)$ and $\hat c_0^{(2)}(k)$ demands a constitutive equation,
the so called closure relation which for any non-trivial case can only
be determined approximatively. Two approaches which proved successful
for the description of fluids with long-range interactions will be
applied here, the hypernetted chain (HNC)~\cite{hansen-mcdonald:86}
and the Rogers-Young (RY) closure relation~\cite{Rogers:84}. They can
both be written as
\begin{equation}\label{eq:hncryclosure}
  h(r) = e^{-\beta u(r)}\left\{1+f(r)^{-1}\left(e^{\chi(r)f(r)}-1\right)\right\}-1\,,
\end{equation}
where $\chi(r)=h(r)-c_0^{(2)}(r)$ is the indirect correlation
function. $f(r)=1-e^{-\xi r}$ is a `mixing function' with an
adjustable parameter $0\leq\xi\leq\infty$ which is either sent to
infinity (HNC)---which is equivalent to letting
$f(r)\rightarrow1$---or chosen to guarantee thermodynamic consistency
between virial and compressibility route to the free energy (RY).

The coupled equations~(\ref{eq:hq}) and (\ref{eq:hncryclosure}) are
iteratively solved by applying the method of fast Fourier transforms
for radially symmetric two-dimensional problems as suggested by
Caillol {\it et al}~\cite{Caillol:81} and as also summarized in
appendix~A of~\cite{Hoffmann:06}. In order to reach rapid convergence
an iteration procedure for the indirect correlation function
$\chi(r)$ is used, since its Fourier transform, $\tilde\chi(k)$,
decays more rapidly with increasing $k$ than $\tilde h(k)$. The
iteration scheme now consists of making an ansatz for $c_0^{(2)}$,
calculating $\chi$ according to equation~(\ref{eq:hq}), obtaining the
next estimate of $c_0^{(2)}$ via equation~(\ref{eq:hncryclosure}),
inserting this into equation~(\ref{eq:hq}), etc., until convergence is
obtained.

Applying this procedure we are able to calculate $\hat
c_0^{(2)}(k;\Gamma)$ for coupling constants $\Gamma$ much larger than
the experimentally known coupling of freezing $\Gamma_f \approx
10$~\cite{Zahn:99} which enables us to calculate the Helmholtz free
energy of the system deep inside the thermodynamically stable
crystalline region.

More accurate pair correlation functions can be obtained from computer
simulations. We have performed extensive Monte Carlo computer
simulations~\cite{allen-tildesley:87} in a quadratic simulation box of
size $L\times L$ comprising $900$ particles employing periodic
boundary conditions in order to measure the pair distribution function
$g_{\rm s}(r)=h_{\rm s}(r)+1$, the subscript `$\rm s$' denoting the
simulation result.  Since the accessible range of $h_{\rm s}(r)$ is
limited to distances $r$ smaller than a cutoff radius $r_c\lesssim
L/2$
we employed an extrapolation technique suggested by
Verlet~\cite{Verlet:68} to obtain the complete pair correlation
function: Verlet defined a closure relation to the Ornstein-Zernicke
equation
\begin{eqnarray} 
h(r)&=h_{\rm s}(r)\qquad &r<r_c\nonumber\\
c_0^{(2)}(r)&=c_{\rm HNC}^{(2)}(r)\qquad &r>r_c\,, \label{eq:Verlet}
\end{eqnarray}
where $c_{\rm HNC}(r)$ is given in equation~(\ref{eq:hncryclosure}).
The Verlet closure relation [equation~(\ref{eq:Verlet})] together with
the Ornstein-Zernicke equation [equation~(\ref{eq:hq})] uniquely
specify the direct correlation function $c_0^{(2)}(r)$ for all radii
$r$ and thus also yield the correlation function in reciprocal space
$\hat c_0^{(2)}(k)$.  As for the HNC and the RY closures the
Ornstein-Zernicke equation and the Verlet closure were solved
iteratively via the indirect correlation function $\chi$.
Furthermore, $r_c$ was chosen the largest root of $h(r)$ still smaller
than $L/2$.

\begin{figure}
\begin{center}
  \includegraphics[width=7.5cm]{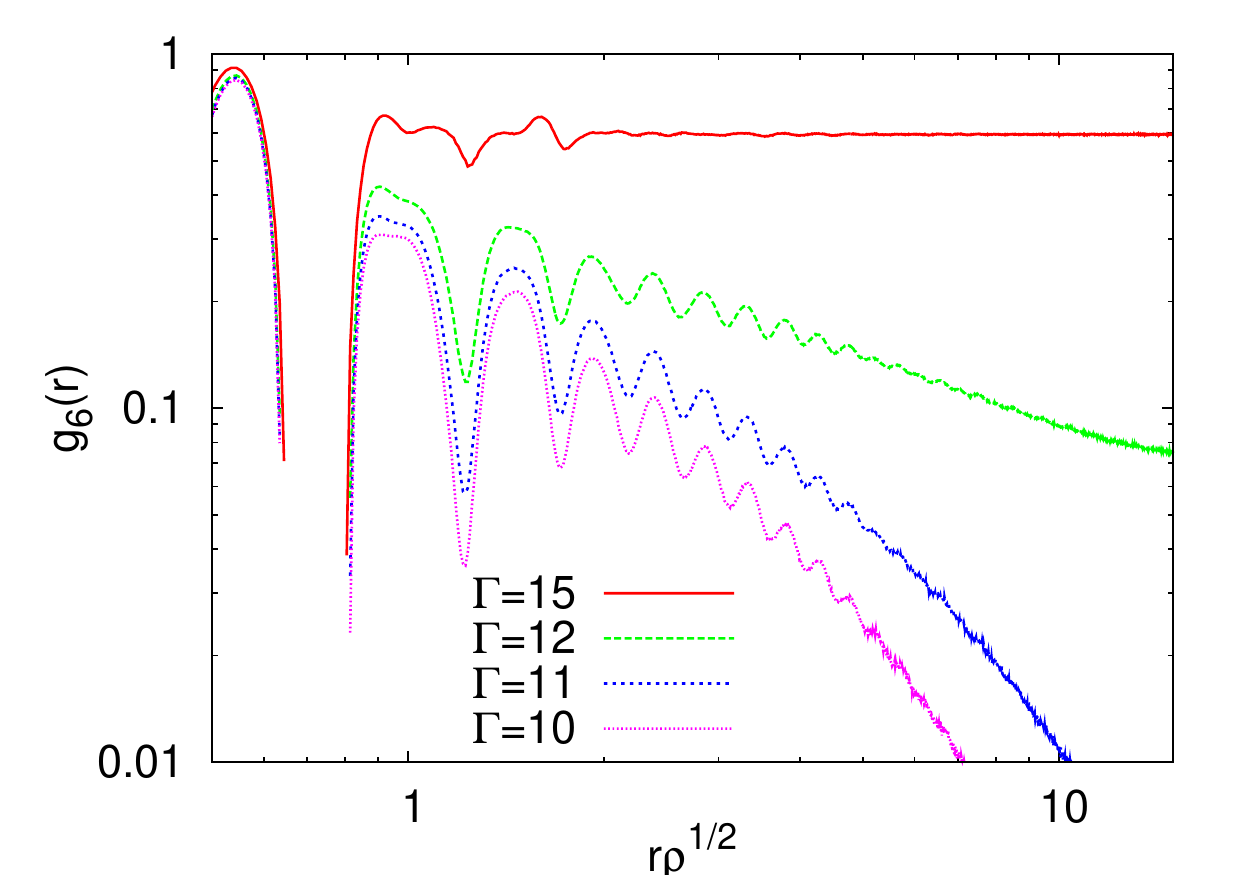}
  \caption{The bond-orientational order parameter $g_6({\bi r})$ for
    different coupling constants $\Gamma$ as obtained by computer
    simulation. $g_6$ decays exponentially for coupling constants
    $\Gamma\lesssim11$ indicating the system to be in the fluid state.}
  \label{fig:g6}
\end{center}
\end{figure}

For the Verlet data we checked that the simulated system does not
crystallize for coupling constants $\Gamma\lesssim11$. Here, the
freezing-criterion was chosen a non-exponential decay of the
bond-orientational order parameter $g_6({\bi
  r})=\left<\exp[i6[\theta({\bi r})-\theta({\bi r}^\prime)]]\right>$,
where $\theta({\bi r})$ is the angle of the bond connecting two
neighboring particles according to the Voronoi construction (see
figure~\ref{fig:g6}).
The application of the Verlet closure within the DFT formalism was thus
restricted to the range $0\leq\hat\Gamma\lesssim11$.

\begin{figure}
\begin{center}
  \includegraphics[width=7.5cm]{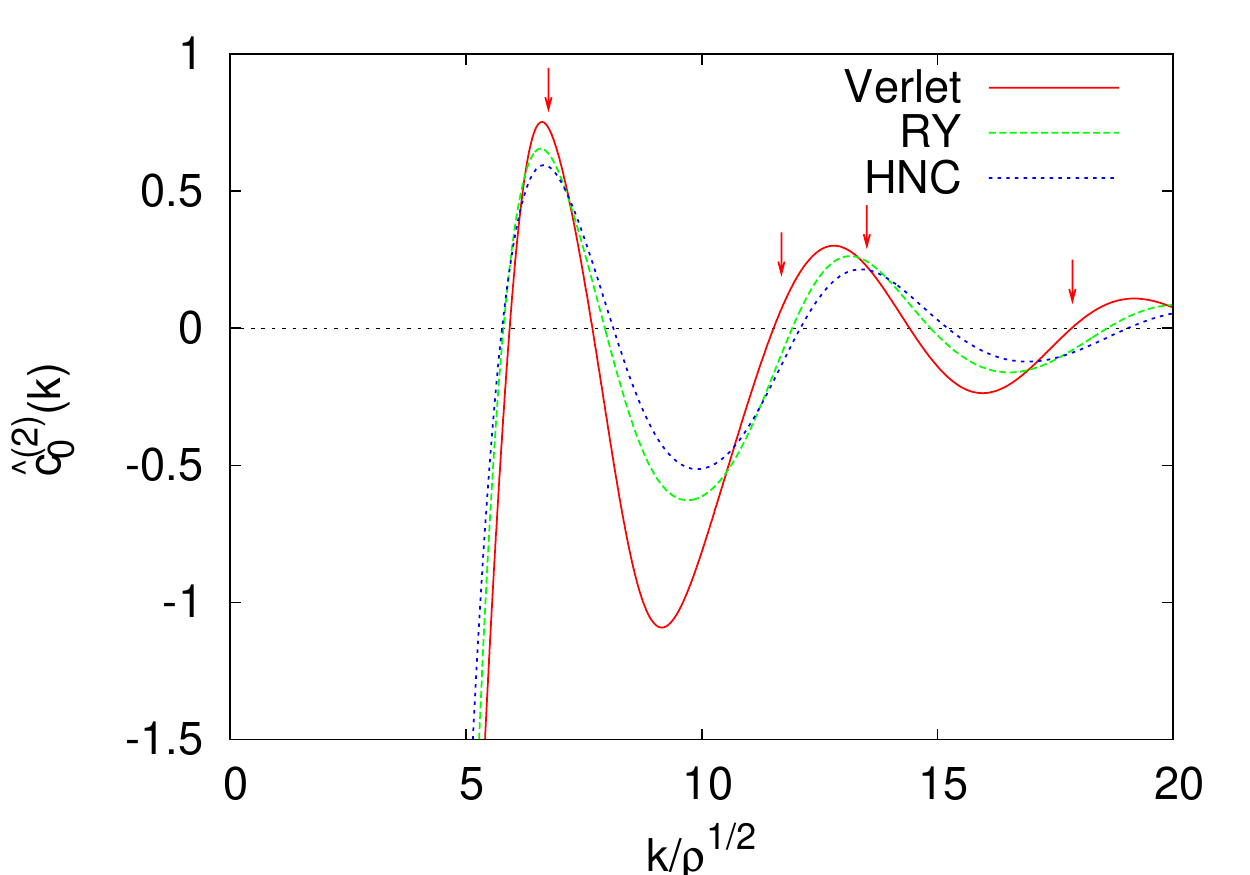}
  \caption{The dimensionless Fourier transform $\hat c_0^{(2)}(k)$ of
    the two-particle direct correlation function at $\Gamma = 9$,
    plotted against $k/ \rho^{1/2}$. Shown are simulation data using
    the Verlet closure, liquid integral equation theory using the RY
    closure, and liquid integral equation theory using the HNC
    closure. The arrows indicate the positions of the first four
    reciprocal lattice vectors of the triangular lattice.}
  \label{fig:c2}
\end{center}\end{figure}

The Fourier transforms $\hat c_0^{(2)}(k)$ of the two-particle direct
correlation functions obtained from the three different closure
relations (HNC, RY, Verlet) are shown in figure~\ref{fig:c2} for
$\Gamma=9$, which is close to the experimentally determined coupling
constant of freezing $\Gamma_f \simeq 10$~\cite{Zahn:99}.
The HNC closure underestimates the pair structure strongly while the
RY closure is closer to the simulation data.  We also show the
positions of the first four reciprocal lattice vectors of the
triangular lattice with lattice constant $a=(\sqrt{3}/2\rho)^{1/2}$.
The value of $\hat{c}^{(2)}_{0}$ at these lattice vectors crucially
influences the solid free energies, as can be seen from
equations~(\ref{eq:fexRYgamma}) and (\ref{eq:gammahatrhohat}).

\begin{figure}
\begin{center}  
\includegraphics[width=7.5cm]{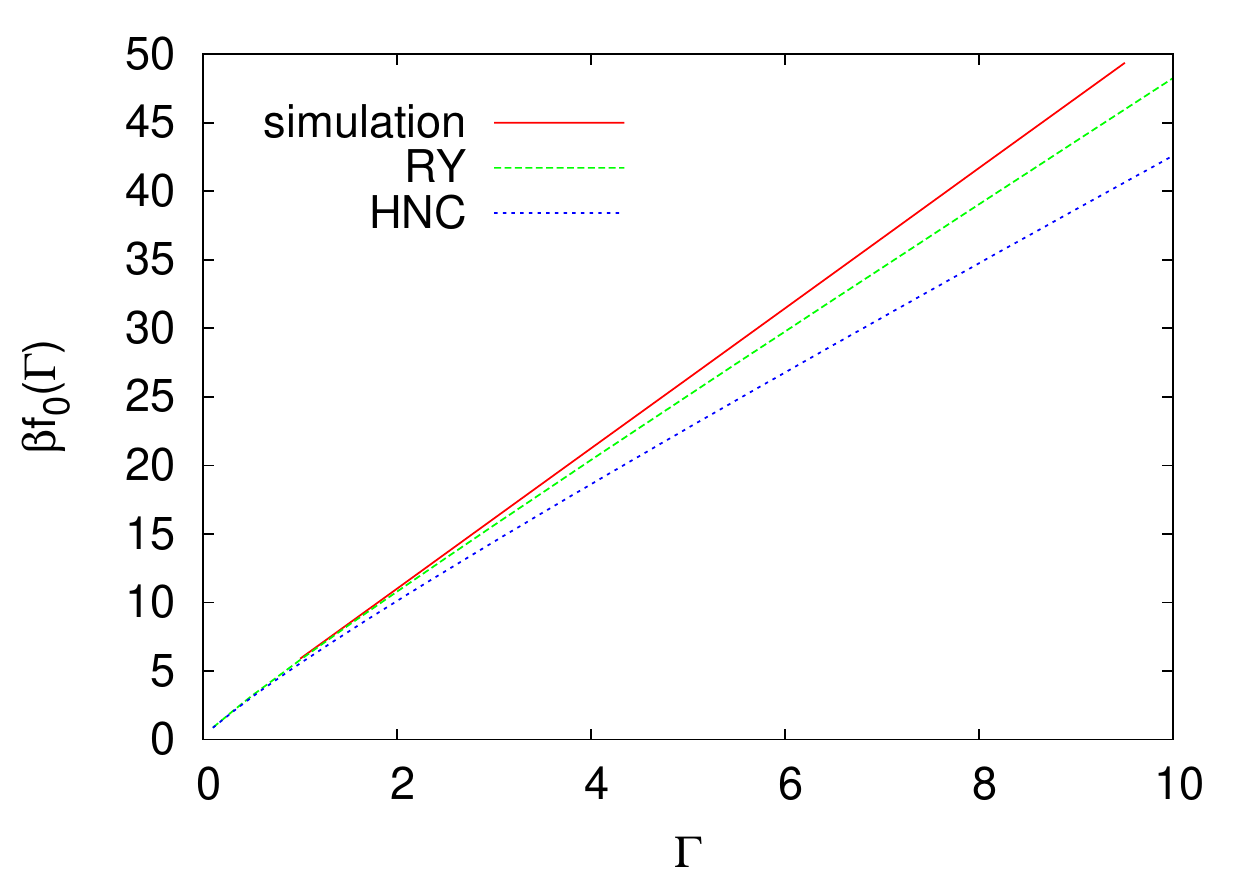}
  \caption{The excess free energy density $\beta f_0(\Gamma)$ as a
    function of coupling constant $\Gamma$ using the Verlet closure,
    the RY closure and the HNC closure.}
  \label{fig:f0}
\end{center}\end{figure}

Within the RY-approach the excess free energy density $f_0$ is
obtained by integrating the compressibility which is inversely
proportional to the static structure factor:
\begin{equation}\label{eq:compressint1}
\beta f_0(\Gamma)=\frac{2}{3}\int_0^\Gamma \frac{\mathrm{d} \Gamma^\prime}{\Gamma^\prime}
\left[\frac{\beta P}{\rho}-1\right]\,,
\end{equation}
where the pressure $P$ is given by
\begin{equation}\label{eq:compressint2}
\frac{\beta P}{\rho}-1=
\frac{2}{3}\int_0^\Gamma\frac{\mathrm{d}\Gamma^\prime}{\Gamma^{\prime1/3}}
\left[1-\hat c_0^{(2)}(k=0;\Gamma^\prime)\right]\,.
\end{equation}
In order to obtain the excess free energy density from the simulation
data we make use of the relation~\cite{Baus:80}
\begin{equation}\label{eq:uex}
\beta \langle  u_{\rm ex}\rangle=
\beta \frac{\partial\beta f_0}{\partial \beta}
=\Gamma\frac{\partial  \beta f_0}{\partial \Gamma}
\end{equation}
between the average excess energy density $\langle u_{\rm ex}
\rangle=\frac{1}{2}\langle \Sigma_{i\neq j}u_{i,j}\rangle$ and $f_0$
and integrate the former.  Note that for both of our approaches, the
RY and the Verlet closure, the virial and the compressibility route
are equivalent. As the energy dominates the free energy in the strong
coupling limit, $\Gamma\gtrsim 1$, the excess free energy density
scales roughly linearly with coupling constant, as can be seen from
figure~\ref{fig:f0}.

For the EMA we need the three-particle correlation function $\tilde
c_0^{(3)}(\bi k,\bi k^\prime; \rho)$ of the underlying fluid for a
wide range of coupling constants.  We use here two conceptually
different approximations: The first approximation is by Denton and
Ashcroft~\cite{Denton:89} (DA) which is based on a weighted density
approximation to the first order direct correlation function
$c^{(1)}(\bi r;\rho(\bi r))$ of an inhomogeneous system. The DA
approach leads to an analytic expression of $\tilde c_0^{(3)}$ in
terms of the one- and two-particle correlation functions $c_0^{(1)}$,
$\tilde c_0^{(2)}$ of the liquid and their derivatives with respect to
density:
\begin{eqnarray}
\fl  \tilde{c}_0^{(3)\rm{DA}}({\bi k}, {\bi k'})=\frac{1}{3}
\Big[
 \tilde{f}^{\rm DA}\left(|{\bi k}|,|{\bi k}'|\right)
+\tilde{f}^{\rm DA}\left(|{\bi k}|,|{\bi k}+{\bi k'}|\right)
+\tilde{f}^{\rm DA}\left(|{\bi k}'|,|{\bi k}+{\bi k'}|\right)\Big],
  \label{eq:c3da}
\end{eqnarray}
where
\begin{eqnarray}
\fl  \tilde{f}^{\rm DA}(k,k')=\frac{1}{c_0^{(1)\prime}}
\left[\tilde c_0^{(2)}(k)\tilde c_0^{(2)\prime}(k')
+\tilde c_0^{(2)\prime}(k)\tilde c_0^{(2)}(k')\right]
-\frac{c_0^{(1)\prime\prime}}{\left[c_0^{(1)\prime}\right]^2}
\tilde c_0^{(2)}(k)\tilde c_0^{(2)}(k').
  \label{eq:c3da2}
\end{eqnarray}
Here, primes denote derivatives with respect to density, as above. The DA
approximation---by construction---fulfills the symmetry condition
\begin{equation}\tilde c_0^{(3)\rm{DA}}(\bi{k}, \bi{k'})=
\tilde c_0^{(3)\rm{DA}}(\bi{k}, \bi{k+k'})=
\tilde c_0^{(3)\rm{DA}}(\bi{k^\prime}, \bi{k+k'})\,.
\end{equation}
The derivatives $\tilde c_0^{(2)\prime}(k)$ were obtained by applying a simple
finite difference method bearing in mind that
\begin{equation}
\fl \rho^2 \tilde c_0^{(2)\prime}(k;\rho)=
 \frac{1}{2}\left[3 \Gamma \frac{\partial\hat c_0^{(2)}(k\rho^{-1/2};\Gamma)}{\partial\Gamma}
- k\rho^{-1/2} \frac{\partial\hat c_0^{(2)}(k\rho^{-1/2};\Gamma)}{\partial k\rho^{-1/2}}-2 \hat c_0^{(2)}(k\rho^{-1/2};\Gamma)\right]\,.
\end{equation}
We calculated $\tilde c_0^{(3)\rm{DA}}(\bi{k}, \bi{k'})$ taking the
direct correlation function from both the RY and the Verlet closure.
As pointed out in Refs.~\cite{Likos:93,Denton:89,Rosenfeld:90b} the DA
model, although itself not derived from a free energy functional but
from an approximate one-particle correlation function, is very
similar to different approaches, all based on taking three successive
functional derivatives of approximate free energy functionals.

We also employed another approximation for $c_0^{(3)}$, namely a
factorization ansatz of Barrat, Hansen and Pastore
(BHP)~\cite{Barrat:88}.  The approximation reads
\begin{equation}\label{cbhp}
  c_{\rm{BHP}}^{(3)}({\bi r},{\bi
    r^\prime})=t(r)t(r^\prime)t(\left|{\bi r}-{\bi r^\prime}\right|)\,.
\end{equation}
The function $t(r)$ can be uniquely determined from the second of the
sum rules in equation~(\ref{eq:sumrules}) which in $r$-space now reads
\begin{equation}\label{eq:sumrulerspace}
\int \mathrm{d} \bi r^\prime c_0^{(3)}(\bi r,\bi r^\prime;\rho)=
\int \mathrm{d} \bi r^\prime t(r)t(r^\prime)t(\left|\bi r-\bi r^\prime\right|)
=\frac{\partial c_0^{(2)}(r;\rho)}{\partial \rho}\,.
\end{equation}
We solved equation~(\ref{eq:sumrulerspace}) numerically for $t(r)$ applying
the method of `steepest descent' as outlined in appendix~B of
reference~\cite{Barrat:88}. As opposed to the simple finite difference
approach above the derivatives $c_0^{(2)\prime}(k)$ were now obtained
by iteratively solving the coupled differentiated Ornstein-Zernicke
equation and the differentiated RY closure relation, as outlined
in appendix B of~\cite{Barrat:88}. Since it proved difficult
to reach convergence of the iteration procedure we did not pursue this
method using the Verlet closure. The triplet-correlation function was
then obtained by a double Fourier transform of
equation~(\ref{eq:sumrulerspace}) using a standard expansion in Legendre
polynomials, as outlined in appendix~A of~\cite{Barrat:88}.

In the single summation in equation~(\ref{eq:gammahatrhohat}) we consider
all RLV's of absolute value $\left|\bi K\right|\leq 33
\left|\bi K_1\right|$, where $\bi K_1$ is the smallest RLV of
the triangular lattice---this comprises the first 299 stars of
RLV's, which is by far sufficient to reach convergence of the single
summation.  The double summation is performed over sets of equivalent
triangles of RLV's which are each characterized by the absolute values
of the two RLV's $\bi K$ and $\bi K^\prime$, and by the
absolute value of there included angle. For the DA model and for the
BHP model we include 42 sets of triangles of RLV's, where that RLV of
the three RLV's, $\bi K$, $\bi K^\prime$, $\bi K-\bi
K^\prime$, with the largest absolute value satisfies $\left|\bi
  K\right|\leq 4 \left|\bi K_1\right|$, which also guarantees
convergence of the double sum.
%
\section{Results}
\label{sec:results}
We first study the influence of the explicit inclusion of the triplet
correlation functions obtained with the DA model and with the BHP
model on the approximate excess free energy according to the TOT as
compared to the simpler SOT, and according to the EMA as compared to
the MWDA, respectively.  For all six approaches we use the two
different closure relations of Rogers and
Young~[equation~(\ref{eq:hncryclosure})], and of
Verlet~[equation~(\ref{eq:Verlet})], respectively.
%
\subsection{Gaussian profiles, no vacancies}
In order to keep things simple in the beginning we keep the number of
particles occupying a lattice site, $n_c$, in
equation~(\ref{eq:ansatzrhoinhomogeneous}) fixed (i.e., $n_c=1$) and
thus end up with a single order parameter, the dimensionless
localization strength $\alpha^*\equiv\alpha/\rho$.
\begin{figure}
\begin{center}
  \includegraphics[width=7.5cm]{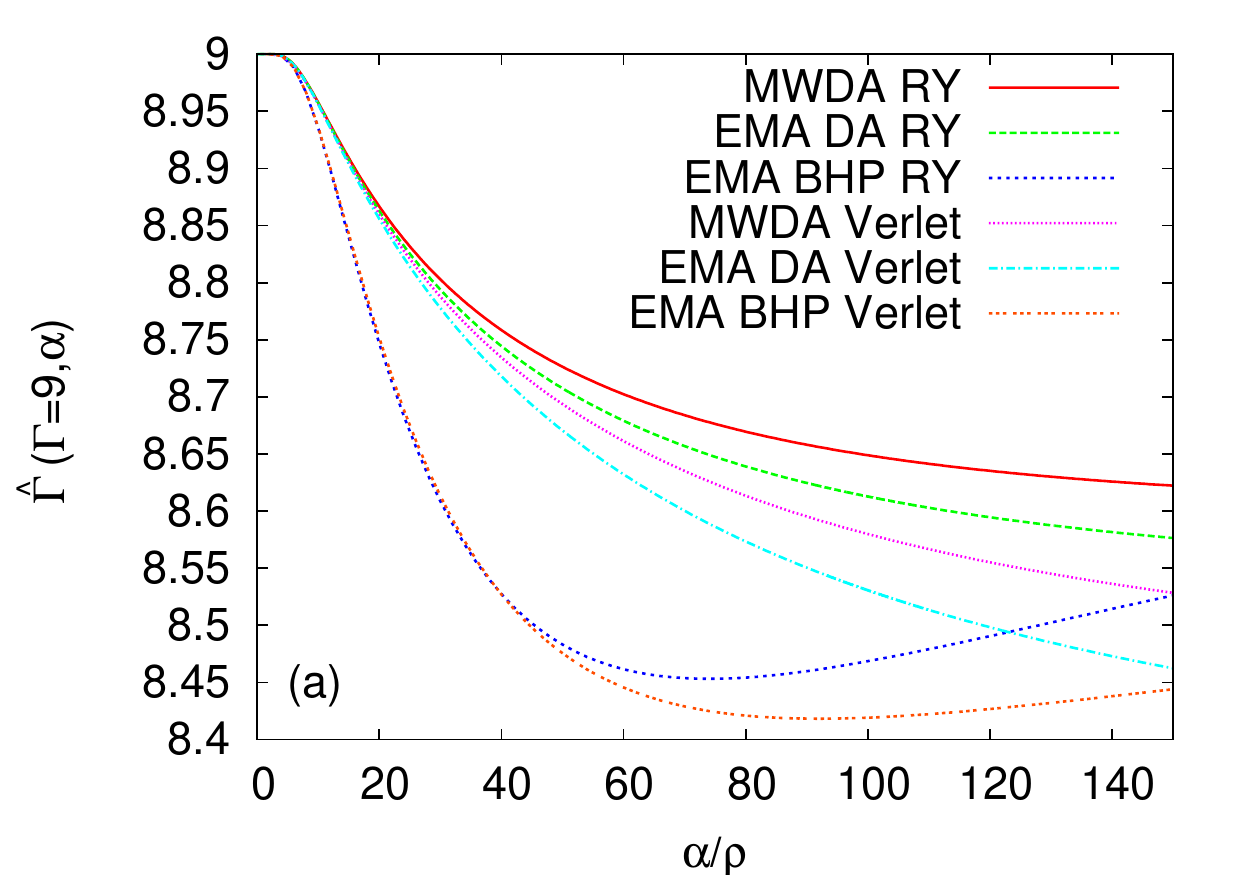}
  \includegraphics[width=7.5cm]{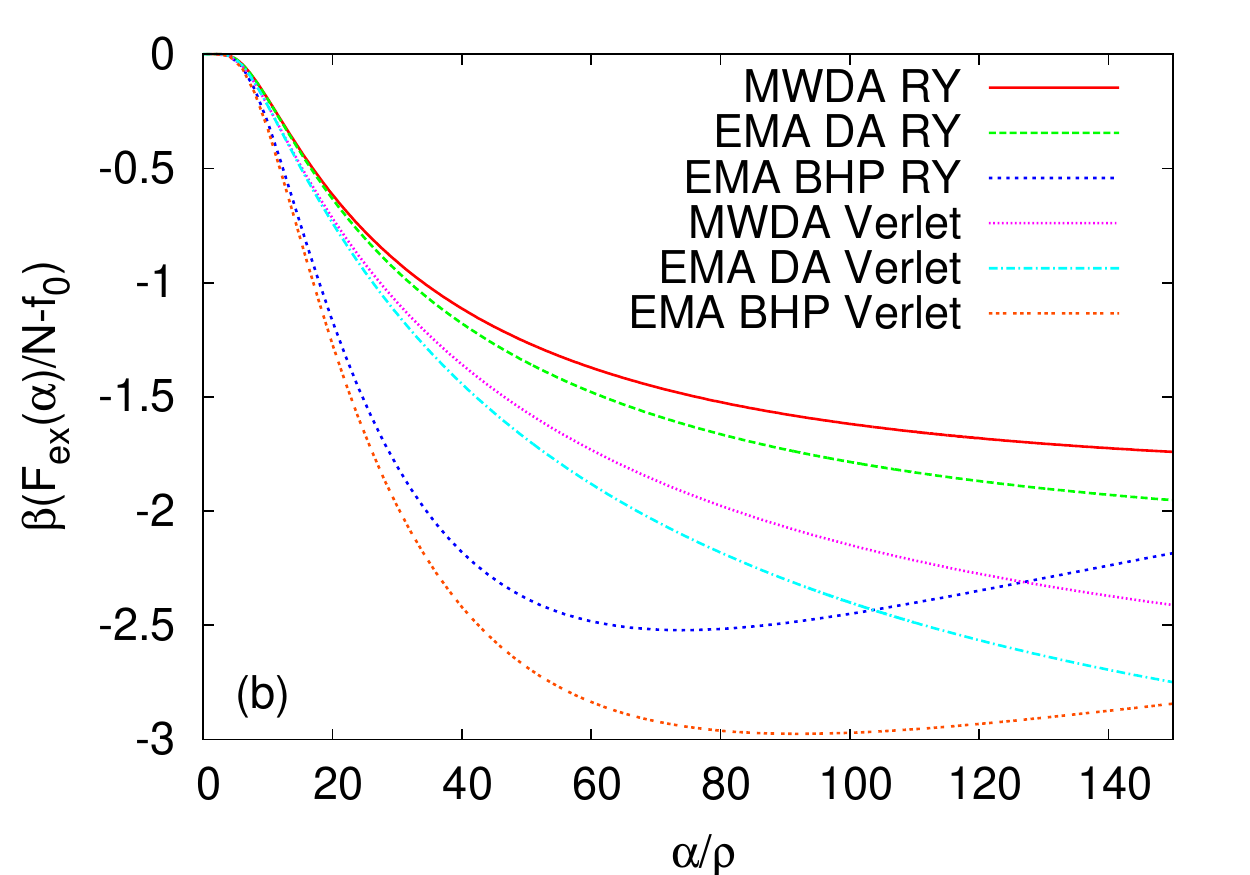}
\\
  \includegraphics[width=7.5cm]{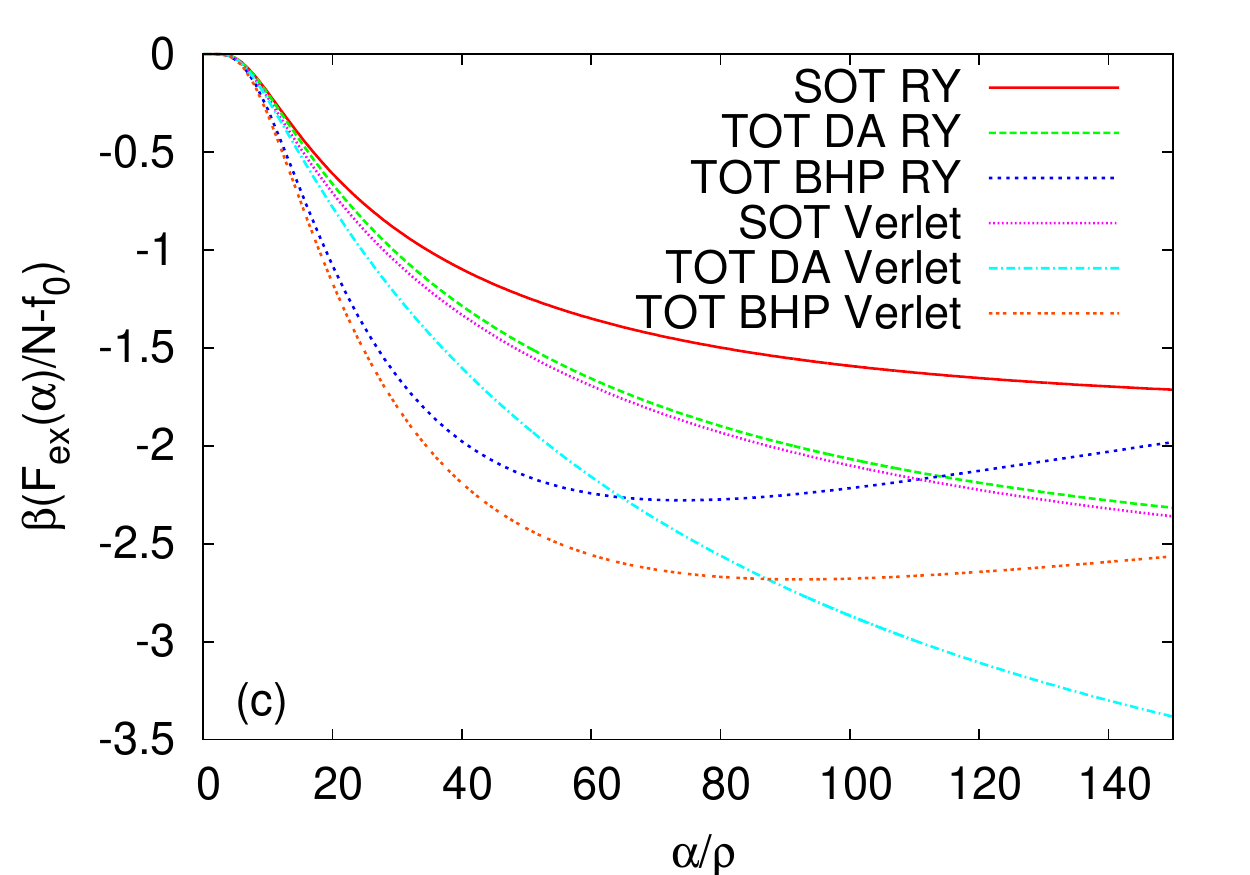}
  \includegraphics[width=7.5cm]{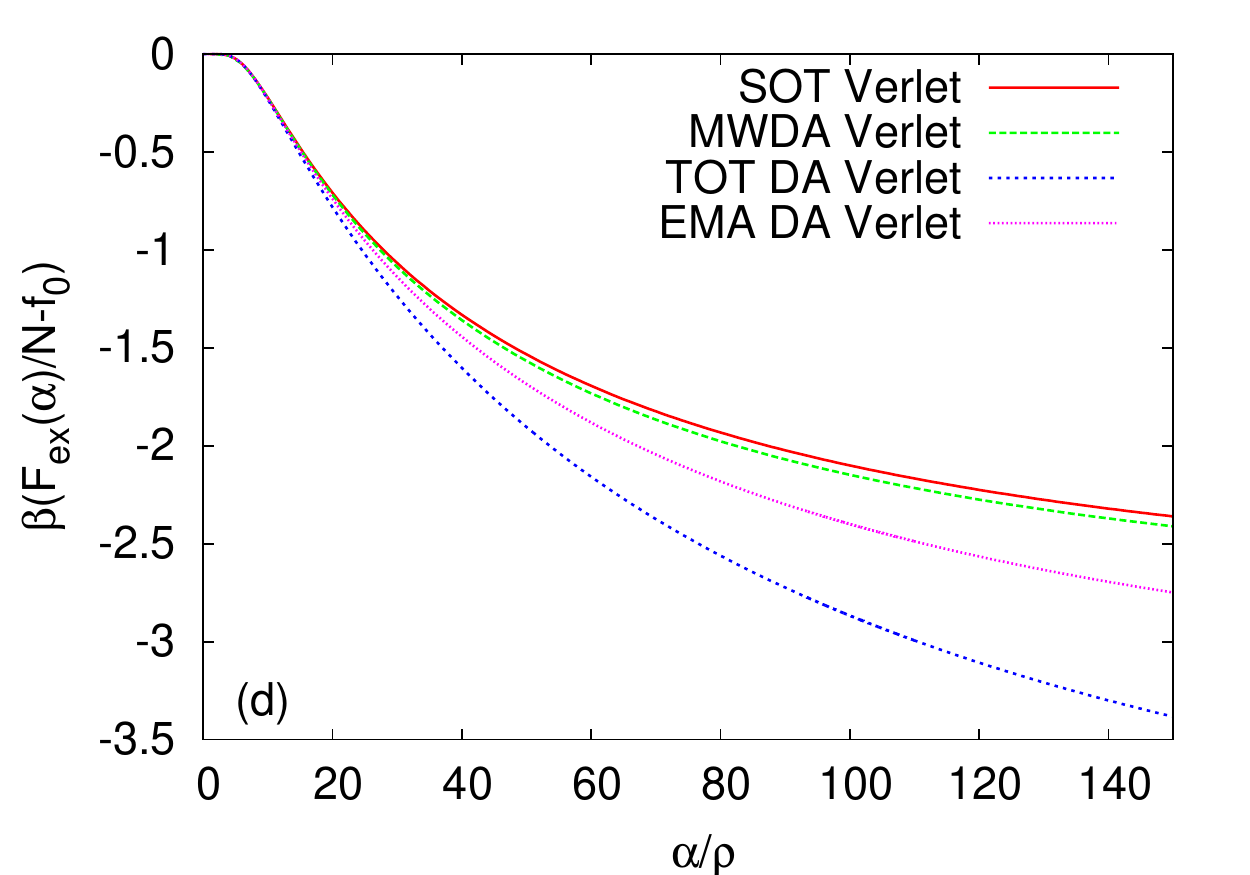}
  \caption{ \label{fig:ghat_fex} (a) The weighted coupling constant as
    a function of $\alpha^*$ within the MWDA and within the EMA
    using $c_0^{(2)}$ from the RY and from the Verlet closure, and
    using $c_0^{(3)}$ from the DA and the BHP model for $\Gamma=9$.
    (b) The approximate excess free energy difference per particle
    $f_0(\hat{\Gamma}(\alpha^*))-f_0(\Gamma)$ as a function of
    $\alpha^*$ for the same approximations as in (a).  (c) The
    approximate excess free energy difference per particle $F_{\rm
      ex}(\alpha^*)/N-f_0(\alpha^*=0)$ as a function of $\alpha^*$
    obtained within the SOT and TOT using the same approximations for
    the two- and three-particle correlation functions as in (a,b) for
    $\Gamma=9$.  (d) Comparison of $F_{\rm ex}$ obtained within the
    four different approximate theories MWDA, EMA, SOT, and TOT using
    $c_0^{(2)}$ from the Verlet closure, and using $c_0^{(3)}$ from the
    DA model for $\Gamma=9$.  }
\end{center}\end{figure}

In figure~\ref{fig:ghat_fex}(a,b) we show the weighted coupling constant
and the associated excess free energy difference per particle between
the solid and the liquid state $F_{\rm
  ex}(\alpha^*)/N-f_0$, according to equation~(\ref{eq:fmwda}), as
functions of localization strength $\alpha^*$ for a value of
$\Gamma=9$ which is close to the experimentally known value of
freezing, $\Gamma_f \simeq 10$~\cite{Zahn:99}, for the MWDA and for
the EMA, using the RY or the Verlet approach to the direct correlation
function and using the two different approaches for the triplet
correlation function, the DA and the BHP model. The latter are both
based on the direct correlation functions used for the respective
two-particle term. In figure~\ref{fig:ghat_fex}(c) the excess free energy
for the simpler SOT and TOT are plotted as a function of $\alpha^*$
for the same approximations to the correlation functions. In
figure~\ref{fig:ghat_fex}(d) the non-perturbative and the perturbative
approaches are compared. Different interesting features of the
different approximations are observed:

(i) For all approaches used except for those where $c_0^{(3)}$ is
obtained within the BHP model the excess free energy decreases
monotonically with increasing localization strength $\alpha^*$,
reaching a plateau for $\alpha^*\approx400$ [c.f.\
figure~\ref{fig:ghat_fex}(b,c)]. However, employing the BHP model to
the triplet-correlation function leads to an increase of
$\hat{\Gamma}(\alpha^*)$ and $F_{\rm ex}(\alpha^*)$ for values of
$\alpha^*\gtrsim80$.  The former behavior is intuitively expected and
has also been observed in the original
MWDA~\cite{Denton:89b}---localization is favoured by the excess part
of the free energy. Once the density peaks become very narrow, a
further increase of $\alpha^*$ does not change the excess free energy
further. On the other hand, the rise of $\hat\Gamma$ and of $F_{\rm
  ex}$ within the BHP model is regarded as unphysical. We therefore do
not consider the BHP model any further.

(ii) Both within the DA model and within the BHP model (for
$\alpha^*\lesssim80$) the sign of the third term in
equation~(\ref{eq:gammahatrhohat}) is negative, i.e., the value of
$\hat\Gamma$ is decreased as compared to the pure MWDA and thus
freezing is favoured. It is also interesting to note, that within
the DA model the triplet-part in equation~(\ref{eq:gammahatrhohat}) is much
smaller than the second-order term while it is significantly larger
within the BHP model. This same behavior had already been found for
hard spheres in three dimensions~\cite{Likos:93}.

(iii) Although the direct correlation functions using the RY- and the
Verlet-closures do not differ by more than $\sim10\%$ at the position
of the most important first RLV (cf.\ figure~\ref{fig:f0}) the
difference in $\hat\Gamma$ between the two is quite pronounced which
is due to the self-consistency relation in
equation~(\ref{eq:gammahatrhohat}).  Furthermore, as shown in
figure~\ref{fig:ghat_fex}(b) the difference in excess free energy is
even more enhanced.

(iv) Inclusion of higher than second-order terms in a non-perturbative
way within the MWDA reduces the excess free energy as compared to the
simpler SOT [c.f.\ figure~\ref{fig:ghat_fex}(d)]. However, inclusion
of higher than third-order terms within the EMA increases the excess
free energy with respect to the TOT.

The total Helmholtz free energy per particle is obtained by adding to
the excess part $F_{\rm ex}$ the ideal part $F_{\rm id}$ according to
equation~(\ref{eq:ftot}).
\begin{figure}
\begin{center}
  \includegraphics[width=7.5cm]{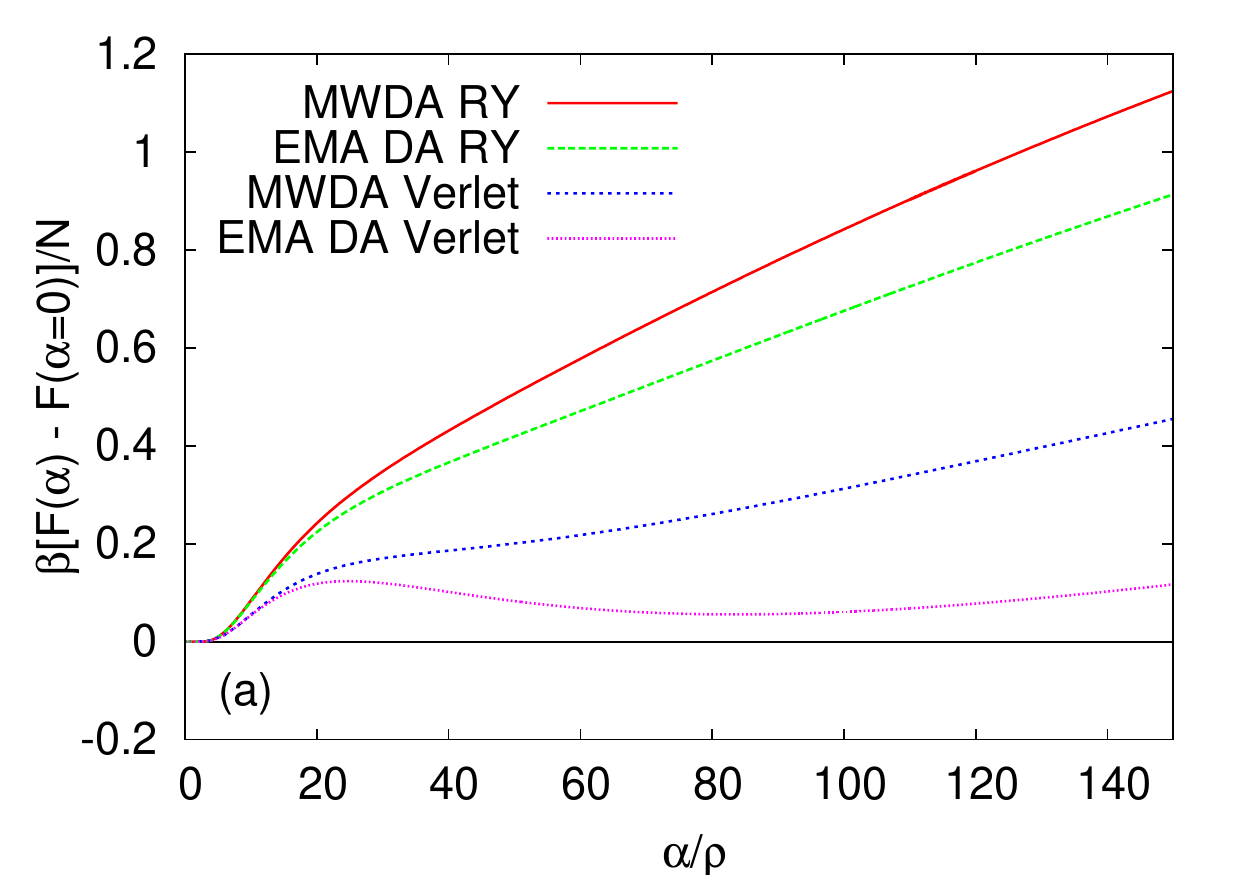}
  \includegraphics[width=7.5cm]{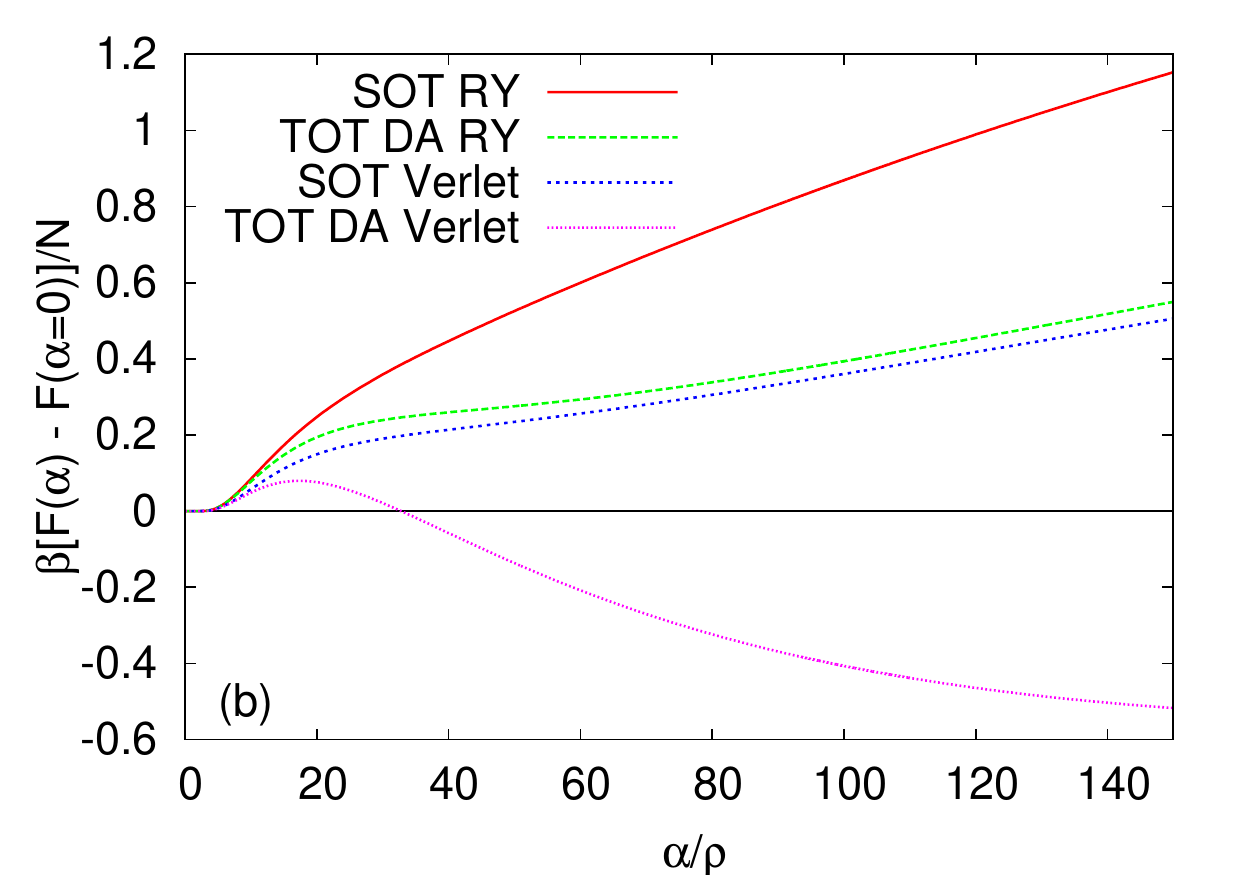}
  \caption{\label{fig:ftot_comp} The total free energy difference per
    particle $\Delta F(\Gamma)/N$ as a function of $\alpha^*$ within
    the MWDA and EMA (a), and within the SOT and TOT (b) using
    $c_0^{(2)}$ from the RY and from the Verlet closure, and using
    $c_0^{(3)}$ from the DA model for $\Gamma=9$.}
\end{center}\end{figure}
The free energy difference per particle $\Delta
F/N=[F(\alpha^*;\Gamma)-F(\alpha^*=0;\Gamma)]/N$ is plotted in
figure~\ref{fig:ftot_comp} as a function of $\alpha^*$, for the same
value of $\Gamma=9$ as in figure~\ref{fig:ghat_fex} for the SOT and
TOT [figure~\ref{fig:ftot_comp}(a)], and for the MWDA and EMA
[figure~\ref{fig:ftot_comp}(b)], respectively, using $c_0^{(2)}$ from
the RY and from the Verlet closure, and using $c_0^{(3)}$ obtained
within the DA model.  It is found that the different curves of $\Delta
F/N$ show qualitatively very different behavior for the coupling of
$\Gamma=9$: While the free energy increases monotonically with
$\alpha^*$ within the SOT and MWDA and within the TOT and EMA
employing the RY closure it displays a local minimum with respect to
$\alpha^*$ at a finite value of $\alpha^*$ within the EMA and TOT,
employing the Verlet closure, this local minimum even turning the deep
global minimum within the TOT at $\alpha^*\approx213$. The appearance
of a global minimum at a finite value of $\alpha^*$ corresponds to a
thermodynamically stable crystalline state while the global minimum at
$\alpha^*=0$ indicates a stable fluid system.

\begin{figure}
\begin{center}
  \includegraphics[width=7.5cm]{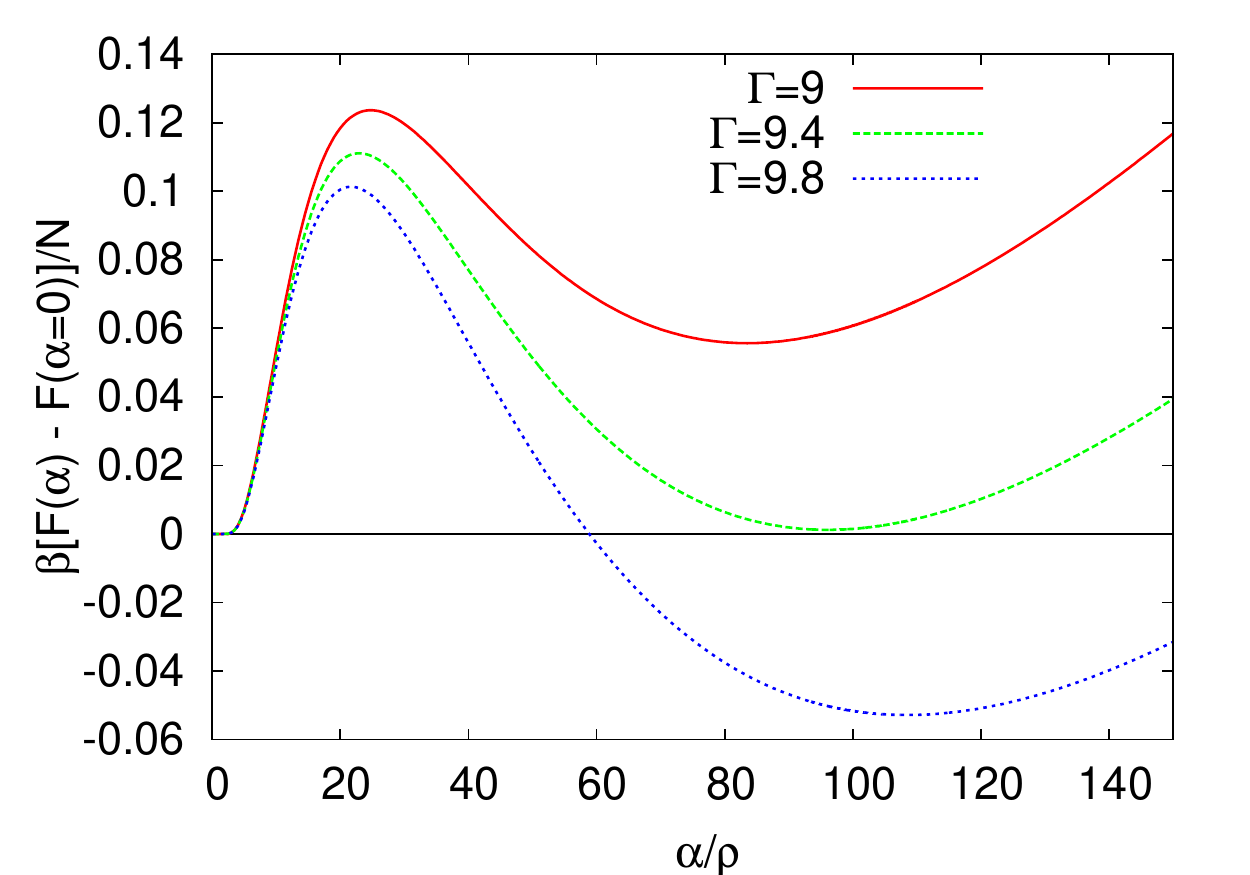}
  \caption{\label{fig:ftot_EMA} The total free energy difference per
    particle $\Delta F(\Gamma)/N$ as a
    function of $\alpha^*$ within the EMA using
    $c_0^{(2)}$ from the Verlet closure, and using 
    $c_0^{(3)}$ from the DA  model for $\Gamma=9,9.4,9.8$.}
\end{center}\end{figure}
In figure~\ref{fig:ftot_EMA}, we display the total free energy
obtained within the EMA employing the DA model with the Verlet closure
for three different values of $\Gamma=9.0,9.4,9.8$. We thus conclude
from figure~\ref{fig:ftot_EMA}---this has already been presented
in Ref.~\cite{Teeffelen:06}---that the EMA employing the Verlet closure and
the DA model yields a transition from the fluid to the solid close to
$\Gamma=9.4$: while for $\Gamma=9.0$ the fluid is stable as indicated
by the minimal value at $\alpha^*=0$, fluid-solid coexistence is
achieved at $\Gamma =9.4$ (see the two equal minima in
figure~\ref{fig:ftot_EMA}). The solid phase, on the other hand, is
clearly stable for $\Gamma=9.8$. The localization parameter at
coexistence is roughly $\alpha^*_{\rm min}\approx100$.

The curves always displays a local minimum with respect to $\alpha^*$
at $\alpha^*=0$. This is in accordance with the mean-field character
of any approximation to the DFT, which ignore fluctuations leading to
a breakdown of long-range order in one and two dimensions.  Therefore, a
first-order transition between fluid and solid state is always
predicted, i.e., the liquid system always has to overcome a free
energy barrier in order to reach the thermodynamically stable
crystalline state.

\begin{figure}
\begin{center}
  \includegraphics[width=7.5cm]{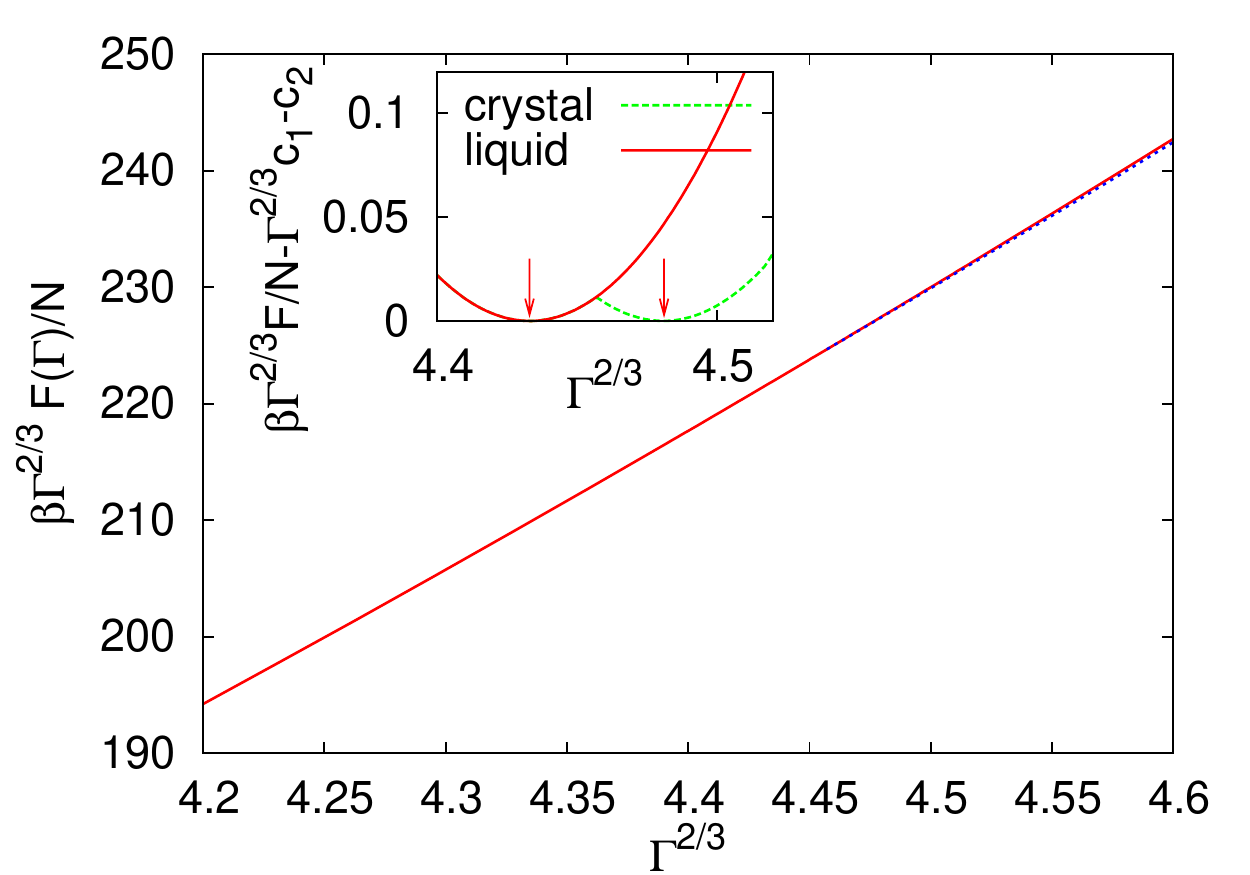}
  \caption{The liquid (solid line) and crystal (dotted line) volume
    free energy densities $\Gamma^{2/3} F/N$ obtained within
    the EMA using the Verlet closure and the DA model as a function of
    $\Gamma^{2/3}$. The inset show the tilted free energy densities
    around the transition values $\Gamma_s$, $\Gamma_f$, as indicated
    by arrows.}
  \label{fig:commontangent}
\end{center}\end{figure}

The freezing and melting transition constants for the first-order
phase transition predicted by the different approximations to the DFT,
$\Gamma_s$ and $\Gamma_f$, respectively, are obtained by using
Maxwell's double tangent construction to the fluid and crystal volume
free energy densities $\Gamma^{2/3} F/N\propto F/V$, where $F$ denotes
the minimum free energy with respect to $\alpha$, and $\Gamma^{2/3}$
is proportional to the average density $\rho$ of the system [c.f.\
equation~(\ref{eq:Gammarho})]. $\Gamma_s$ and $\Gamma_f$ correspond to the
freezing and melting densities, $\rho_s$ and $\rho_f$, respectively.
The volume free energy density is exemplarily shown for the EMA using
the Verlet closure and the DA model in figure~\ref{fig:commontangent}.
Within this approximation we obtain freezing and melting with a narrow
coexistence gap $\Delta \Gamma = \Gamma_s -\Gamma_f$. 
\begin{table}
  \caption{
    Freezing and melting parameters $\Gamma_f$ and $\Gamma_s$, the 
    widths of the coexistence regions $\Delta\Gamma=\Gamma_s-\Gamma_f$, the 
    relative displacement parameters $\gamma$, and the pressures $P$ at coexistence 
    obtained within: the SOT with the RY closure (first row); the TOT with the RY 
    closure (second row); the TOT with the Verlet closure (third row); 
    the MWDA with the RY closure (forth row); the EMA with the RY 
    closure (fifth row); the EMA with the Verlet closure (sixth row), 
    where all three-particle correlation functions were obtained with the DA model 
    using the respective pair-correlation function as input. The last row displays 
    experimental parameters for the isotropic-hexatic transition,    
    the hexatic-crystal transition and the Lindemann parameter, 
    obtained from real-space microscopy measurements 
    of magnetic colloids confined to an air-water interface. }
\label{tab:freezing}
\begin{center}\begin{tabular}{l|llllll}
    & $\Gamma_f$ & $\Gamma_s$ & $\Delta\Gamma$& $\gamma$& $\beta P(\Gamma_f)/\rho$\\
    \hline
    SOT  with RY                & 42.85   & 42.92  & 0.07   & 0.017& 288.3\\
    TOT  with RY                & 13.49   & 13.62  & 0.13   & 0.021& 93.1\\
    TOT  with Verlet            & 6.79    & 6.97   & 0.18   & 0.019& 53.1\\
    MWDA  with RY               & 41.07   & 41.13  & 0.06   & 0.017& 276.1\\
    EMA   with RY               & 23.0    & 23.08  & 0.09   & 0.020& 156.9\\
    EMA   with Verlet           & 9.33    & 9.49   & 0.16   & 0.020& 72.6\\
    Experiment                  &    10.0  & 10.75  & -     & 0.038& - \\
\end{tabular}\end{center}\end{table}
Table~I summarizes the freezing/melting parameters for all the
approximations made. The data are compared against experimental
results obtained from real-space microscopy measurements of magnetic
colloids confined to an air-water interface. The experiments give
freezing with an intermediate hexatic phase. The liquid-solid
transition has also been studied using numerical
simulation~\cite{Haghgooie:05,Loewen:96} yielding a slightly higher
inverse transition temperature between $12.0$ and $12.25$ but these
investigations suffer from finite size effects.

As becomes evident from Table~I, the SOT, TOT, and MWDA are not
quantitatively satisfying theories as they either over- or
underestimate the freezing coupling.  Note that the overestimation of
the freezing coupling within SOT and MWDA are the reason why it is not
possible to feed the ``exact'' pair structure into these theories. At
such high coupling, no fluid pair structures are available since the
fluid spontaneously crystallizes in the simulation.  The EMA, on the
other hand, yields results in close agreement with experimental data.
The TOT obviously underestimates the freezing coupling by a factor of
$\approx 2$.

More detailed, structural information can be extracted from the
localization parameter of the coexisting solid.  For all
approximations used we find localization parameters at freezing in the
range $ 99 < \alpha^*_{\rm{min}}(\Gamma_s) < 115$.  Strictly
speaking, the localization parameter has no counterpart in ``real'' 2D
systems since the particles are not localized due to long range
fluctuations.  However, if one relates the particle displacements to
that of their nearest neighbor, one can define a finite quantity as
$\gamma=\rho\left\langle \left({\bi u}_i - {\bi
      u}_{i+1}\right)^2\right\rangle$, where ${\bi u}_i$ and ${\bi
  u}_{i+1}$ are the displacement vectors of neighboring lattice sites.
Disregarding nearest-neighbor correlations $\left\langle {\bi u}_i
  \cdot {\bi u}_{i+1}\right\rangle$, $\gamma$ can be estimated. Since
the nearest-neighbor correlations $\left\langle {\bi u}_i \cdot {\bi
    u}_{i+1}\right\rangle$ are expected to be positive:
\begin{equation}
  \label{eq:lindemannratio}
  \gamma\lesssim 2 \rho \left\langle {\bi u}_i^2\right\rangle
  \approx 2/\alpha^*_{\rm{min}}.
\end{equation}
By this relation, the localization parameter of the coexisting solid
gives a prediction for $\gamma$ which is included in Table~I.  From
experiments, $\gamma$ is known to be close to $\cong
0.038$~\cite{Zahn:99}. This was shown to be in accordance with
harmonic lattice theory~\cite{Froltsov:05}.  The EMA yields
$\gamma\lesssim 0.020$, i.e.  the EMA roughly overestimates the
localization of the particles by a factor of $2$. $\gamma$ is {\it
  smaller} than the experimental value, contrarily to what was
expected from the inequality~(\ref{eq:lindemannratio}).  This shows
that there is still a need to improve the theories in order to
correctly predict localization properties. A similar overestimation of
the localization is also common in weighted density approximations in
three spatial dimensions~\cite{Denton:89b}.

Another quantity of interest, which is directly connected to the
Helmholtz free energy is the pressure at coexistence which is also
included in Table~I. It is obtained via the
equations~(\ref{eq:compressint1}), (\ref{eq:compressint2}), (\ref{eq:uex}),
depending on whether the RY closure or the simulation data were used.
%
\subsection{Gaussian profiles, allowing for vacancies}
In this subsection, we relax the constraint of zero vacancy
concentration, $1-n_c=0$, in equation~(\ref{eq:ansatzrhoinhomogeneous}) and
instead minimize the total free energy with respect to the two
parameters $\alpha$ and $n_c$, respectively. However, instead of
calculating the phase diagram for all approximations to the DFT and to
the pair- and triplet-correlation functions, we focus here on the two
non-perturbative approaches, the MWDA using the RY closure and the EMA
using the Verlet closure and the DA model.
\begin{figure}
\begin{center}
  \includegraphics[width=7.5cm]{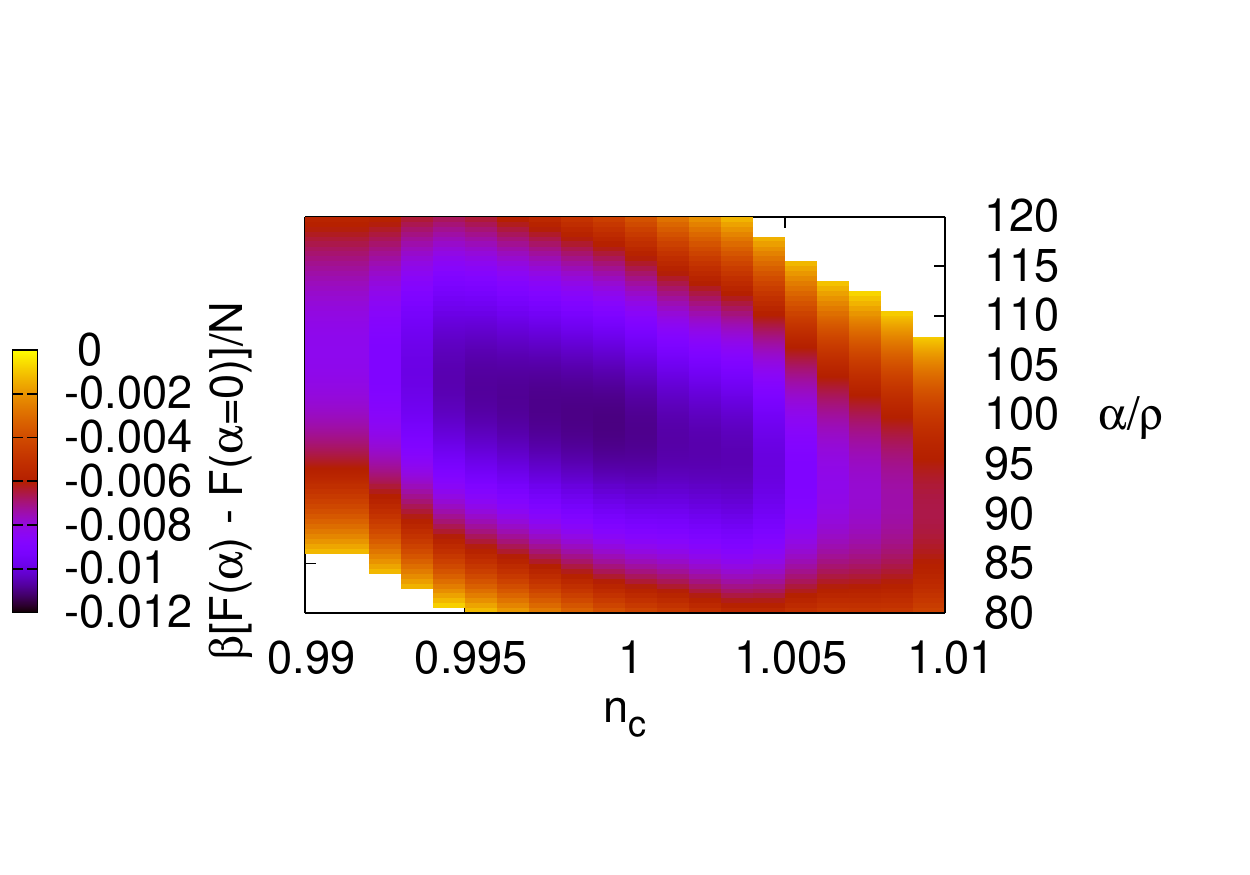}
  \includegraphics[width=7.5cm]{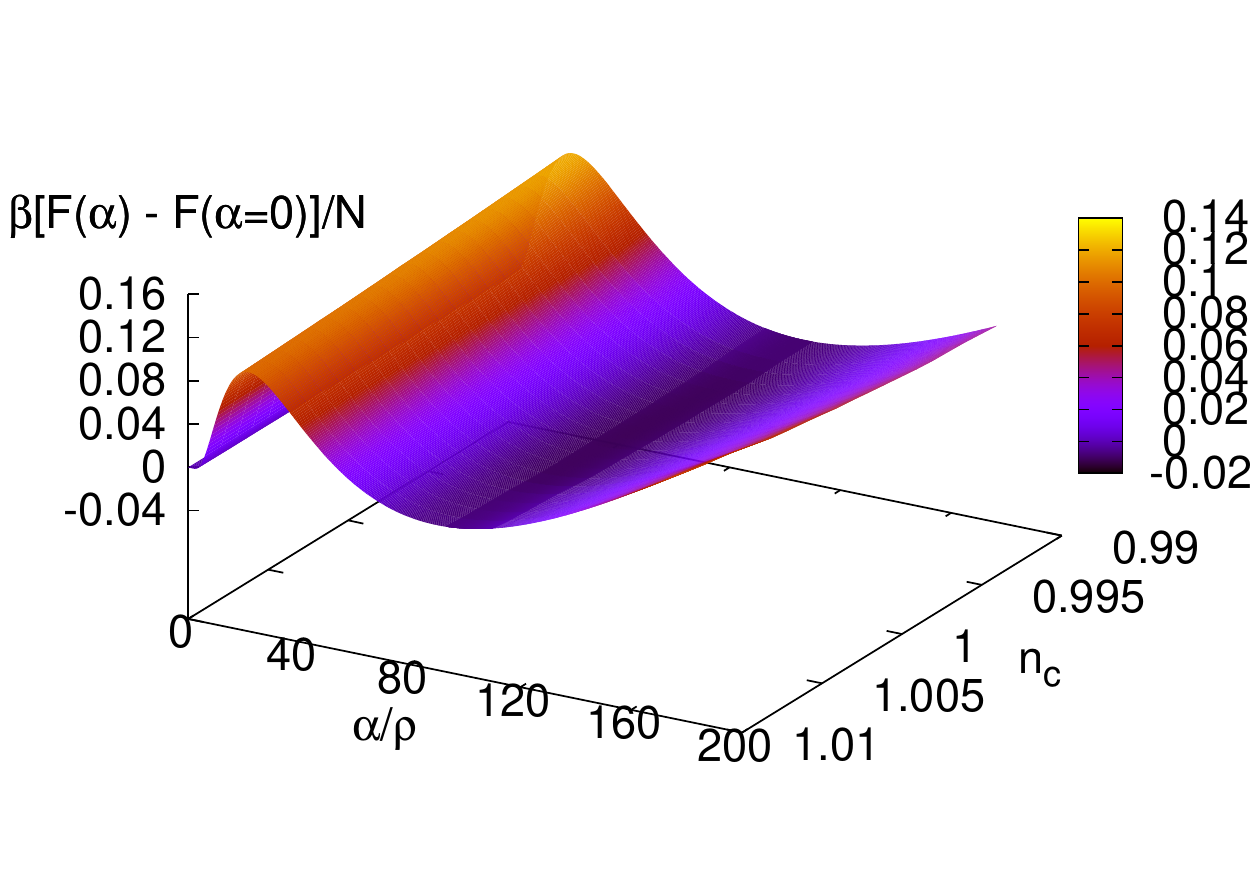}
  \caption{\label{fig:ftot_EMA_nc_alpha} The total free energy
    difference per particle $\Delta F(\Gamma)/N$ as a function of
    $\alpha^*$ and $n_c$ within the EMA using $c_0^{(2)}$ from the
    Verlet closure, and using $c_0^{(3)}$ from the DA model for
    $\Gamma=9.49$. The left panel displays a zoom-in of the right
    panel.}
\end{center}\end{figure}
In figure~\ref{fig:ftot_EMA_nc_alpha} we plot the approximate total
free energy per particle of the EMA as a function of $\alpha$ and
$n_c$ for the freezing coupling constant obtained at fixed $n_c=1$,
$\Gamma=9.49$.  The minimum of the total free energy is slightly
shifted in $n_c$ and $\alpha$ from ($n_c\approx1$,
$\alpha^*\approx98.7$) towards ($n_c\approx0.998$,
$\alpha^*\approx100.5$). As can be seen in
figure~\ref{fig:ftot_EMA_nc_alpha}(a), the difference in total free
energy per particle between the two configurations is only of the
order $10^{-4}k_BT$, which has no influence on the phase diagram
within the accuracy given in Table~I.

For the simpler MWDA, however, the vacancy concentration is
substantially larger, which has pronounced effects on the phase
diagram. In particular, we find the coupling constants of freezing and
melting reduced to ($\Gamma_f\approx37.35, \Gamma_s\approx37.45$), the
liquid being in coexistence with the triangular crystal at the
parameters $n_c\approx0.966$, $\alpha^*\approx200.5$, i.e., the
relaxation of $n_c$ improves the prediction of the freezing coupling
while the Lindemann parameter $\gamma\approx0.01$ is by a factor of
$\approx2/3$ smaller than predicted within the simpler theory keeping
$n_c=1$ fixed which---compared to the experiment---is worse than the
result from the constrained theory.
\subsection{Free minimization}
\label{sec:freemin}
In this final subsection we completely remove the constraint of
Gaussian density peaks. Instead, we minimize the density functional
with respect to a free, periodic density field $\rho(x,y)$, which has
the periodicity of the hexagonal lattice with lattice constant
$a=(\sqrt{3}n_c/2\rho)^{1/2}$, as above. As laid out in
Ref.~\cite{SvenPhilMag}, we minimize the density functional of the SOT
with the RY closure with respect to $\rho(\bi r)$ by calculating the
overdamped relaxation dynamics of a highly ordered hexagonal crystal
with the help of dynamical
DFT~\cite{Marconi:99,Dzubiella:03,Archer:04,SvenPRL} according to
\begin{equation}\label{eq:ddft}
\frac{\partial\rho(\vec r,t)}{\partial t}= \beta D  \vec\nabla\cdot\left(\rho(\vec r,t)\vec\nabla
\frac{\delta F[\rho(\vec r,t)]}{\delta \rho(\vec r,t)} \right)\,,
\end{equation}
where $\beta D$ is the mobility coefficient, which sets the Brownian
time scale $\tau_B=(\rho D)^{-1}$. Since in this work we are only
interested in the equilibrium state reached after long time, $\tau_B$
is irrelevant in the following considerations, i.e., we use
equation~(\ref{eq:ddft}) just as a minimization procedure to the static
DFT.  Starting from an initial density profile $\rho({\bi r},t=0)$,
equation~(\ref{eq:ddft}) is solved numerically for times
$(t/\tau_B)\lesssim10$ applying a finite difference method and keeping
the coupling constant $\Gamma$ fixed.  The maximum time is chosen
large enough to guarantee convergence towards a (local) minimum of the
free energy landscape. The rectangular periodic box of size $L_x\times
L_y=\sqrt{3}a\times a$ with a discretization of $256\times 128$
lattice points comprises $2n_c$ particles. Due to lattice symmetry, it
suffices to solve the problem in a single elementary cell. For
$\rho({\bi r},t=0)$ we choose a superposition of sharply localized
Gaussians according to equation~(\ref{eq:ansatzrhoinhomogeneous}) with a
large localization strength of $\alpha^*=200$. 

At first, we fix $n_c=1$ and calculate the equilibrium density
profiles and the according approximate Helmholtz free energies for
various coupling constants $0<\Gamma\leq 62.5$. 
\begin{figure}
\begin{center}
  \includegraphics[width=7.5cm]{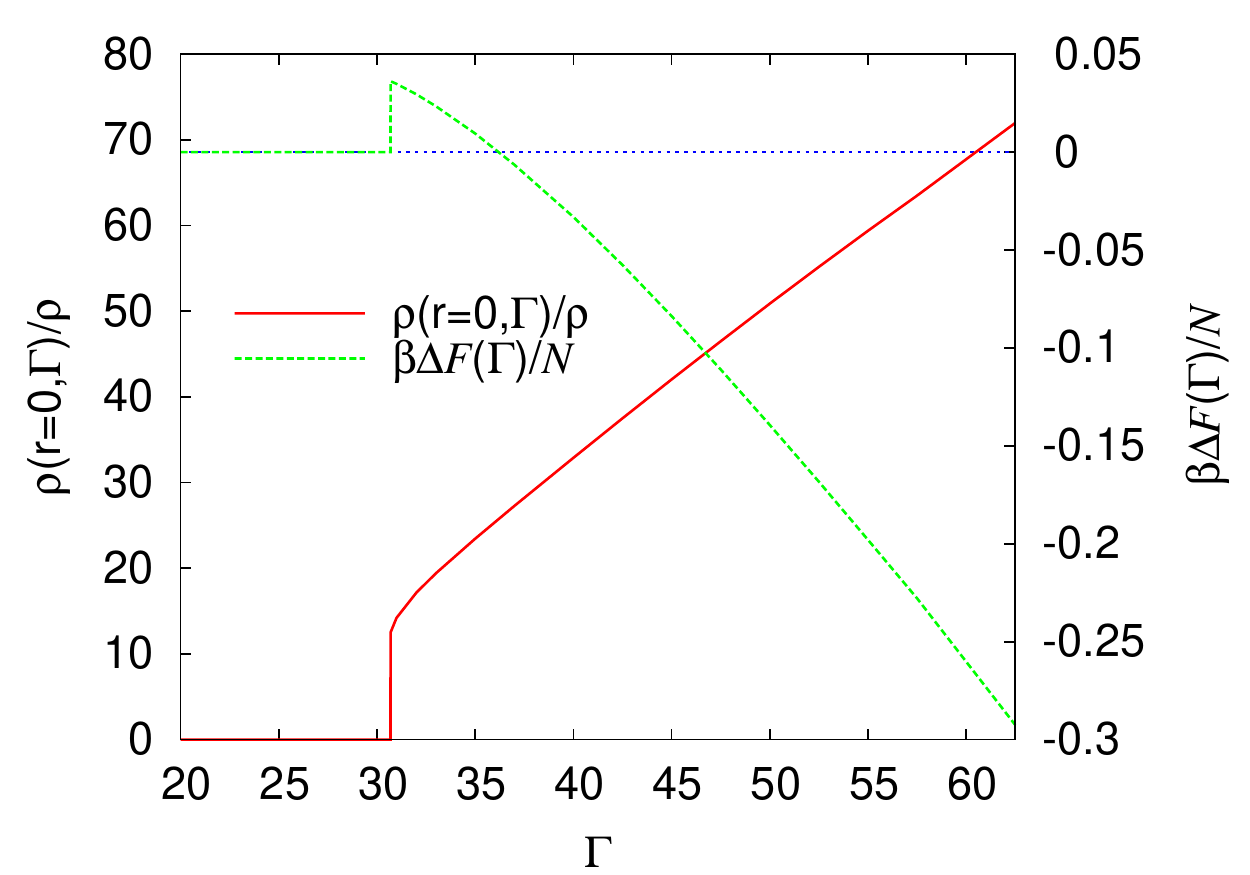}
  \caption{\label{fig:adiabatic} Height of the density peak $\rho({\bi
      r},\Gamma)$ and difference in free energy
    $\Delta F(\Gamma)$ as a function of $\Gamma$ obtained from
    dynamical DFT using the SOT with the RY closure.}
\end{center}\end{figure}
In figure~\ref{fig:adiabatic} we plot the difference in Helmholtz free
energy density $\Delta F(\Gamma)/N=F[\rho({\bi
  r},t\rightarrow\infty;\Gamma)]/N-f_0(\Gamma)$ between the final
(solid/liquid) and the liquid state as a function of $\Gamma$. The
system remains crystalline for couplings $\Gamma\gtrsim 30.7$.
However, the free energy difference is negative only for
$\Gamma\gtrsim 36.2$, which is equivalent with thermodynamic
stability. As for the Gaussian parametrization coexistence is found in
a narrow gap around $\Gamma\approx 36.2$ which we do not specify
here.

\begin{figure}
\begin{center}
  \includegraphics[width=7.5cm]{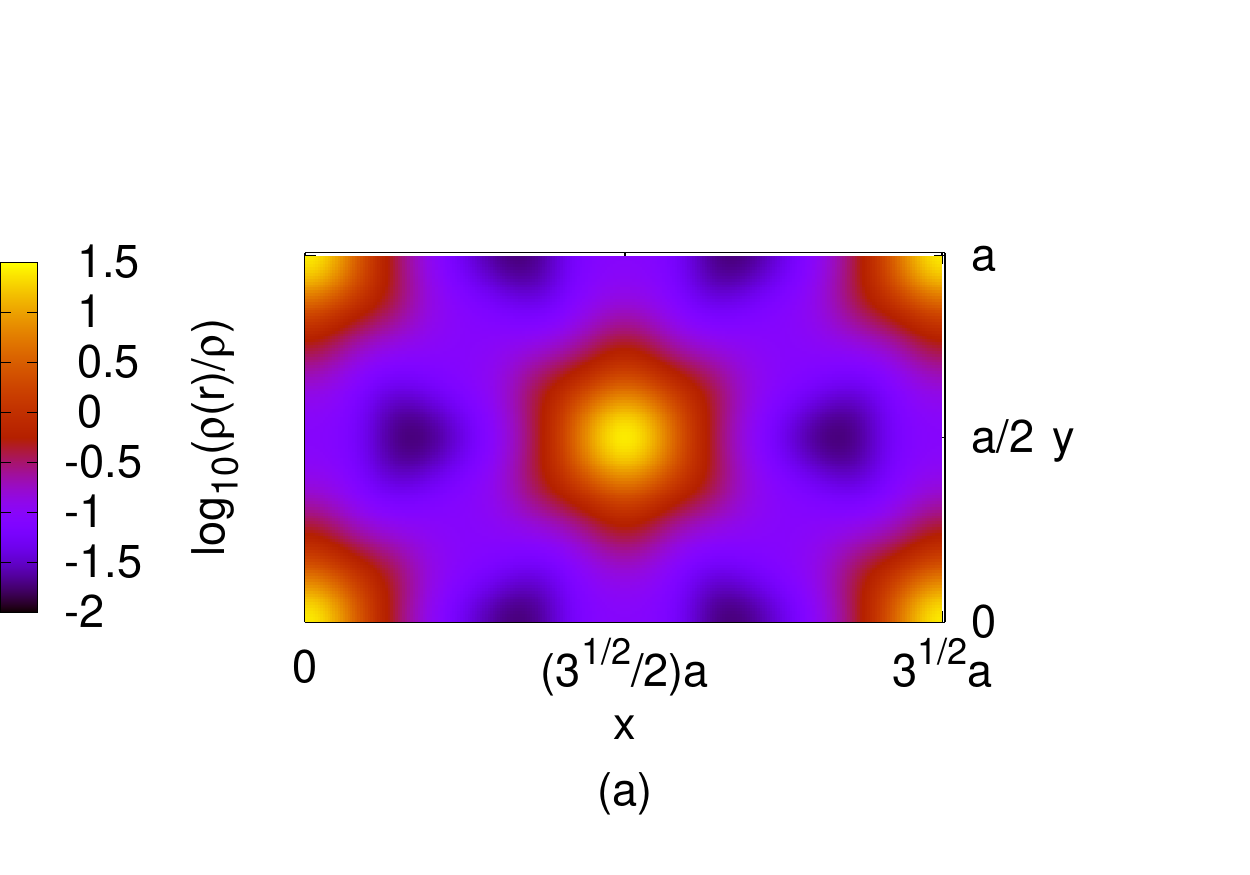}
  \includegraphics[width=7.5cm]{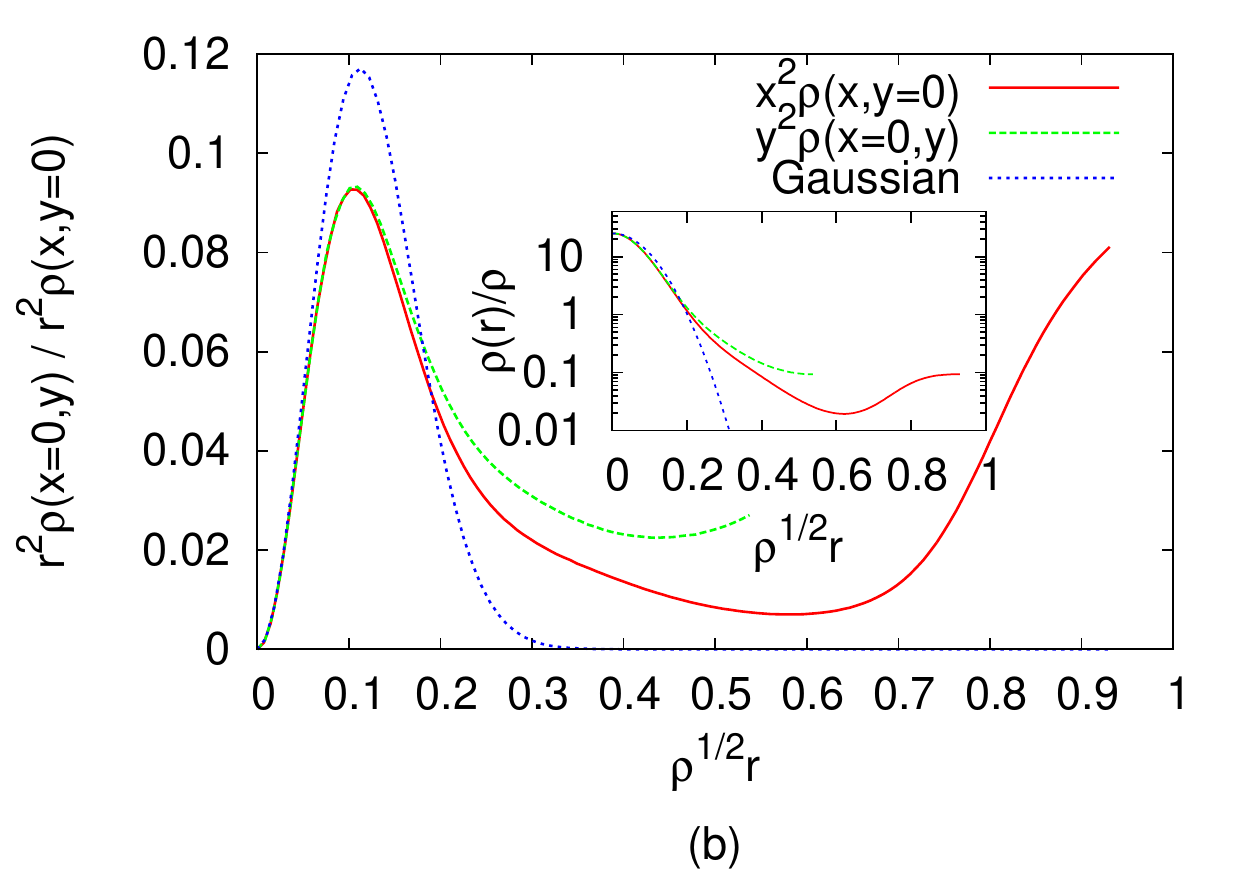}
  \caption{\label{fig:rhor_xy} (a) The density profile $\rho({\bi r})$
    obtained from dynamical DFT using the SOT with the RY closure for
    $\Gamma=36$ which is close to freezing. (b) The quantity
    $r^2\rho({\bi r};t\rightarrow\infty)$ along the straight line
    connecting two nearest neighbours
    [$y^2\rho(x=0,y;t\rightarrow\infty)$, i.e., in the [10]-direction]
    and along the line connecting two next-nearest neighbours
    [$x^2\rho(x,y=0;t\rightarrow\infty)$, i.e., in the
    [11]-direction], both drawn from the center to the respective edge
    of the box in (a). The two curves are compared to a Gaussian of
    the same amplitude at ${\bi r}=0$. The inset displays the bare
    density along the same lines and the bare Gaussian.}
\end{center}\end{figure}
In figure~\ref{fig:rhor_xy}(a) we plot the equilibrium density profile
$\rho(\bi r;\Gamma)$ for $\Gamma=36$ which is close to freezing. In
figure~\ref{fig:rhor_xy}(b) the quantity $r^2\rho({\bi r})$, where $r$
is the distance from a lattice vector, is shown along the two
directions [11] and [10], corresponding to cuts through the density
plane in figure~\ref{fig:rhor_xy}(a) along the $x$- and the $y$- axis,
respectively, which is compared to a Gaussian of the same height as
the density peaks. It is found that the density profile has an
isotropic Gaussian form for small distances from the origin $r\lesssim
0.1/\rho^{1/2}$. For larger distances, however, i.e., where the
density is of the order $\rho({\bi r})\lesssim\rho$, the density
profile significantly deviates from a Gaussian form. In particular, we
observe the establishment of ``bridges'' of higher density between
neighbouring lattice sites, whereas the density is significantly lower
between next-nearest neighbours. This counter-intuitive behavior was
also found applying the MWDA to hard sphere crystals in three spatial
dimensions~\cite{Ohnesorge:93}. However, computer simulations revealed
that the behavior should be the opposite. Although we did not measure
the density profiles of the two-dimensional dipolar system in computer
simulations, we expect a similar behavior: The probability density
should be enhanced along the [11]-direction as compared to the
[10]-direction.

\begin{figure}
\begin{center}
  \includegraphics[width=7.5cm]{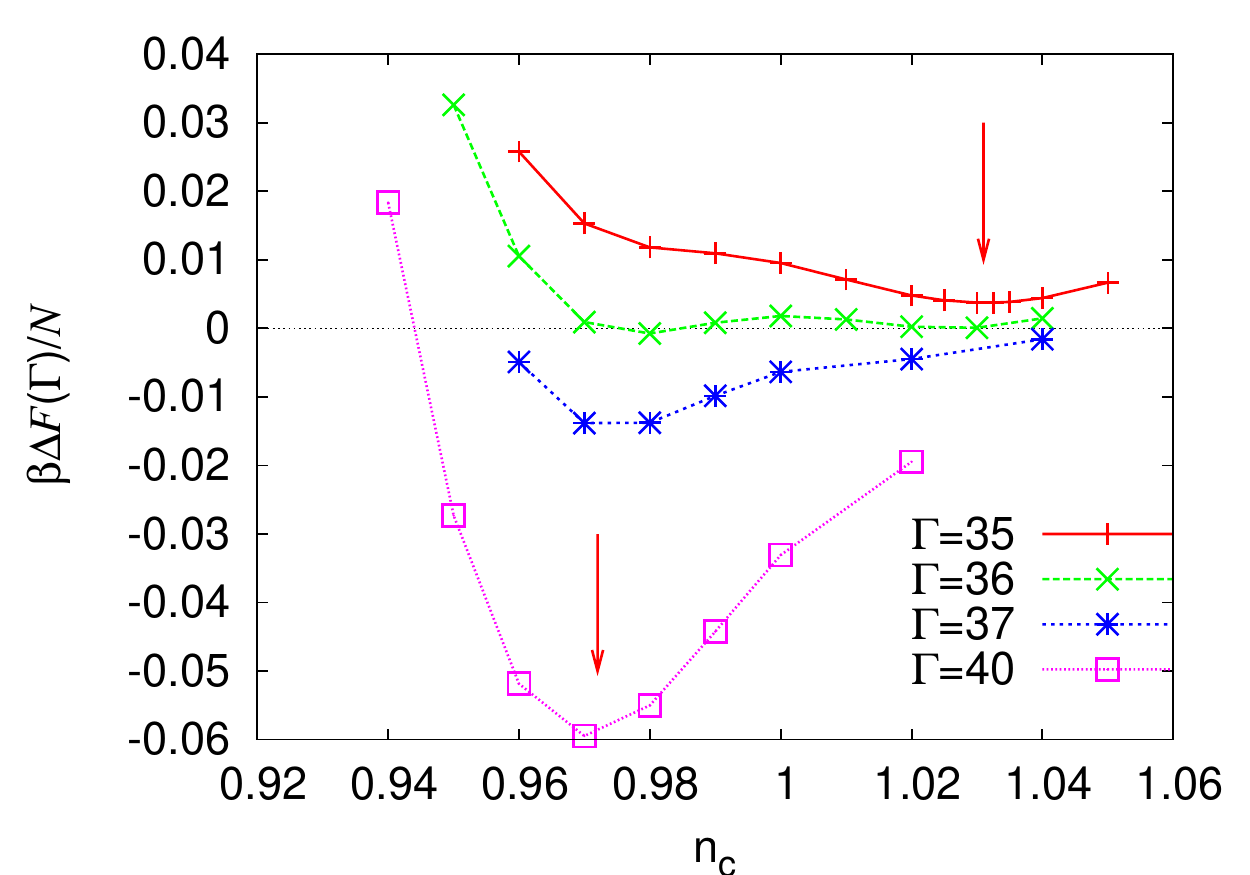}
  \caption{\label{fig:DeltaF_nc} The difference in Helmholtz free
    energy per particle $\Delta F(n_c;\Gamma)/N$ as a function of
    $n_c$ for different coupling constants $\Gamma=35,37,40$. The
    arrows indicate the positions of the minima.}
\end{center}\end{figure}

We also performed the minimization procedure for different
vacancy concentrations. In figure~\ref{fig:DeltaF_nc} we show the free energy
difference $\Delta F(n_c;\Gamma)$ as a function of $n_c$ for four
different values of $\Gamma$.
We find, that for crystals in equilibrium, i.e., for $\Gamma\gtrsim
36$, the equilibrium vacancy concentration is $1-n_c\approx0.03$.
However, the overheated crystal which is metastable for
$31\lesssim\Gamma\lesssim36$ prefers a vacancy concentration of
$1-n_c\approx-0.03$, implying interstitials instead of vacancies.  We
note that most of the point defects in the experimental realization of
the dipolar system appear in pairs or in pairs of pairs as
dislocations or pairs of dislocations, respectively~\cite{Zahn:99}.
%
\section{Discussion and concluding remarks}
In conclusion, we have demonstrated that density functional theory is
able to quantitatively predict the freezing transition of a
two-dimensional colloidal system with long-range $1/r^3$-interactions
in good agreement with experimental and simulation data. In complete
analogy to systems in 3D, the appearance of long-range interactions
requires the explicit inclusion of three-particle correlation
functions of the liquid in the construction of the weighted
density~\cite{Likos:92,Likos:93}.  Furthermore, the predicted
transition temperatures are very sensitive towards slight changes of
the two- and three-particle correlation functions of the underlying
fluid. A highly accurate input of the same is therefore crucial.

The obtained density functional can be used in future studies in order
to approach more complicated situations such as crystals in
confinement~\cite{Ohnesorge:93}, under gravity~\cite{Biben:94}, and
crystal-fluid interfaces~\cite{Ohnesorge:93,Marr:93}. By extending the
static functional to Brownian
dynamics~\cite{SvenPhilMag,Marconi:99,Dzubiella:03,Archer:04,SvenPRL},
one may even address nonequilibrium situations.  One possible problem
to tackle is heterogeneous nucleation upon temperature quenches and
subsequent crystal growth as outlined recently in Ref.~\cite{SvenPRL}.
\ack We thank Norman Hoffmann for helpful discussions.  This work has
been supported by the DFG within the Collaborative Research Centre
SFB~TR6, ``Physics of Colloidal Dispersions in External Fields'',
project section C3.
%


\section*{References}
\begin{thebibliography}{10}

\bibitem{Evans:79}
Evans R.
\newblock {\em Adv. Phys.}, 28:143, 1979.

\bibitem{Oxtoby:91}
Oxtoby~D W.
\newblock In J~P Hansen, Levesque D, and Zinn-Justin J, editors, {\em Liquids,
  Freezing and the Glass Transition}, volume Session LI (1989) of {\em Les
  Houches Summer Schools of Theoretical Physics}, page 147, Amsterdam, 1991.
  North Holland.

\bibitem{Singh:91}
Singh Y.
\newblock {\em Phys.~Rep.}, 207:351, 1991.

\bibitem{Loewen:94}
L{\"o}wen H.
\newblock {\em Phys.~Rep.}, 237:249, 1994.

\bibitem{Loewen:02}
L{\"o}wen H.
\newblock {\em J.~Phys.:~Condens.~Matter}, 14:11897, 2002.

\bibitem{Likos:01}
Likos~C N.
\newblock {\em Phys. Rep.}, 348:267--349, 2001.

\bibitem{Rosenfeld:89}
Rosenfeld Y.
\newblock {\em Phys.~Rev.~Lett.}, 63:980, 1989.

\bibitem{Rosenfeld:97}
Rosenfeld Y, Schmidt M, L{\"o}wen H, and Tarazona P.
\newblock {\em Phys.~Rev.~E}, 55:4245, 1997.

\bibitem{Schmidt:03}
Schmidt M.
\newblock {\em J.~Phys.:~Condens.~Matter}, 15:S101, 2003.

\bibitem{Curtin:88}
Curtin~W A.
\newblock {\em J.~Chem.~Phys.}, 88:7050, 1988.

\bibitem{Curtin:85}
Curtin~W A and Ashcroft~N W.
\newblock {\em Phys.~Rev.~A}, 32:2909, 1985.

\bibitem{deKuijper:90}
de~Kuijper~A, Vos~W L, Barrat J-L, Hansen J-P, and Schouten~J A.
\newblock {\em J.~Chem.~Phys.}, 93:5187, 1990.

\bibitem{Likos:92}
Likos~C N and Ashcroft~N W.
\newblock {\em Phys.~Rev.~Lett.}, 69:316, 1992.

\bibitem{Zeng:90}
Zeng~X C and Oxtoby~D W.
\newblock {\em J.~Chem.~Phys.}, 93:2692, 1990.

\bibitem{Rosenfeld:90}
Rosenfeld Y.
\newblock {\em Phys.~Rev.~A}, 42:5978, 1990.

\bibitem{Tejero:93}
Tejero~C F and Cuesta~J A.
\newblock {\em Phys.~Rev.~E}, 47:490, 1993.

\bibitem{Teeffelen:06}
van Teeffelen~S, Likos~C N, Hoffmann N, and L{\"o}wen H.
\newblock {\em Europhys.~Lett.}, 75:583, 2006.

\bibitem{Ohnesorge:93}
Ohnesorge R, L{\"o}wen H, and Wagner H.
\newblock {\em Europhys.~Lett.}, 22:245, 1993.

\bibitem{Ramakrishnan:79}
Ramakrishnan~T V and Yussouff M.
\newblock {\em Phys.~Rev.~B}, 19:2775, 1979.

\bibitem{Laird:90}
Laird~B B and Kroll~D M.
\newblock {\em Phys.~Rev.~A}, 42:4810, 1990.

\bibitem{Denton:89b}
Denton~A R and Ashcroft~N W.
\newblock {\em Phys.~Rev.~A}, 39:4701, 1989.

\bibitem{Likos:93}
Likos~C N and Ashcroft~N W.
\newblock {\em J.~Chem.~Phys.}, 99:9090, 1993.

\bibitem{Barrat:88}
Barrat J-L, Hansen J-P, and Pastore G.
\newblock {\em Mol.~Phys.}, 63:747, 1988.

\bibitem{Denton:89}
Denton~A R and Ashcroft~N W.
\newblock {\em Phys.~Rev.~A}, 39:426, 1989.

\bibitem{Barrat:88b}
Barrat J-L, Xu~H, Hansen J-P, and Baus M.
\newblock {\em J.~Phys.~C.}, 21:3165, 1988.

\bibitem{Zahn:99}
Zahn K, Lenke R, and Maret G.
\newblock {\em Phys.~Rev.~Lett.}, 82:2721, 1999.

\bibitem{Lin:06}
Lin~S Z, Zheng B, and Trimper S.
\newblock {\em Phys.~Rev.~E}, 73:066106, 2006.

\bibitem{Haghgooie:05}
Haghgooie R and Doyle~P S.
\newblock {\em Phys.~Rev.~E}, 72:011405, 2005.

\bibitem{Loewen:96}
L{\"o}wen H.
\newblock {\em Phys.~Rev.~E}, 53:R29, 1996.

\bibitem{Froltsov:03}
Froltsov~V A, Blaak R, Likos~C N, and L{\"o}wen H.
\newblock {\em Phys.~Rev.~E}, 68:061406, 2003.

\bibitem{Zahn:97}
Zahn K, M\'endez-Alcaraz~J M, and Maret G.
\newblock {\em Phys.~Rev.~Lett.}, 79:175, 1997.

\bibitem{Weeks:81}
Weeks~J D.
\newblock {\em Phys.~Rev.~B}, 24:1530, 1981.

\bibitem{Hoffmann:06}
Hoffmann N, Likos~C N, and L{\"o}wen H.
\newblock {\em J.~Phys.:~Condens.~Matter}, 18:10193, 2006.

\bibitem{hansen-mcdonald:86}
Hansen J-P and McDonald~I R.
\newblock {\em Theory of simple liquids}.
\newblock Academic Press, London, 2nd edition, 1986.

\bibitem{Rogers:84}
Rogers~F J and Young~D A.
\newblock {\em Phys.~Rev.~A}, 30:999, 1984.

\bibitem{Caillol:81}
Caillol~J M, Levesque D, and Weis J-J.
\newblock {\em Mol.~Phys.}, 44:733, 1981.

\bibitem{allen-tildesley:87}
Allen~M P and Tildesley~D J.
\newblock {\em Computer Simulation of Liquids}.
\newblock Clarendon Press, Oxford, 1987.

\bibitem{Verlet:68}
Verlet L.
\newblock {\em Phys.~Rev.}, 165:201, 1968.

\bibitem{Baus:80}
Baus M and Hansen J-P.
\newblock {\em Phys.~Rep.}, 59:1, 1980.

\bibitem{Rosenfeld:90b}
Rosenfeld Y, Levesque D, and Weis J-J.
\newblock {\em J.~Chem.~Phys.}, 92:6818, 1990.

\bibitem{Froltsov:05}
Froltsov~V A, Likos~C N, L{\"o}wen H, Eisenmann C, Gasser U, Keim P, and Maret
  G.
\newblock {\em Phys.~Rev.~E}, 71:031404, 2005.

\bibitem{SvenPhilMag}
L{\"o}wen H, Likos~C N, Assoud L, Blaak R, and van Teeffelen~S.
\newblock {\em Philos.~Mag.~Lett.}, 87:847, 2007.

\bibitem{Marconi:99}
Marconi U~M B and Tarazona P.
\newblock {\em J.~Chem.~Phys.}, 110:8032, 1999.

\bibitem{Dzubiella:03}
Dzubiella J and Likos~C N.
\newblock {\em J.~Phys.:~Condens.~Matter}, 15:L147, 2003.

\bibitem{Archer:04}
Archer~A J and Evans R.
\newblock {\em J.~Chem.~Phys.}, 121:4246, 2004.

\bibitem{SvenPRL}
van Teeffelen~S, Likos~C N, and L{\"o}wen H.
\newblock {\em Phys.~Rev.~Lett.}, 100:108302, 2008.

\bibitem{Biben:94}
Biben T, Ohnesorge R, and L{\"o}wen H.
\newblock {\em Europhys.~Lett.}, 28:665, 1994.

\bibitem{Marr:93}
Marr~D W and Gast~A P.
\newblock {\em Phys.~Rev.~E}, 47:1212, 1993.

\end{thebibliography}
\end{document}